\documentclass[onecolumn,pre,amsmath,amssymb,showpacs,superscriptaddress]{revtex4}
\usepackage{graphics}
\usepackage{graphicx}
\usepackage{subfigure}
\usepackage{psfrag}
\usepackage{color}
\usepackage{bm}
\usepackage{natbib}

\def\x{{\bm x}}
\def\bxi{{\bm \xi}}
\def\d{{\bm d}}
\def\c{{\bm c}}

\def\u{{\bm u}}

\def\q{{\bm q}}
\def\bxi{{\bm{\xi}}}

\def\erfc{{\mbox{erfc}}}
\def\tPr{\tiny{\mbox{Pr}}}

\def\dpart#1#2{{\dfrac{\partial #1}{\partial #2}}}
\def\be{\begin{equation}}
\def\ee{\end{equation}}

\begin{document}
\title{Discrete unified gas kinetic scheme for all Knudsen number flows: II. Compressible case}
  \author{Zhaoli Guo}
 \email[Email:]{zlguo@mail.hust.edu.cn}
 \affiliation{State Key Laboratory of Coal Combustion, Huazhong University of Science and Technology, Wuhan 430074, China}
 \affiliation{Department of Mathematics, Hong Kong University of Science and Technology, Clear Water Bay, Hong Kong, China}
 \author{Ruijie Wang}
 \email[Email:]{ruijie.wang@ust.hk}
 \affiliation{Nano Science and Technology Program, Hong Kong University of Science and Technology, Clear Water Bay, Hong Kong, China}
 \author{Kun Xu}
 \email[Email:]{makxu@ust.hk}
 \affiliation{Department of Mathematics, Hong Kong University of Science and Technology, Clear Water Bay, Hong Kong, China}

\begin{abstract}
This paper is a continuation of our earlier work [Z.L. Guo {\it et al.}, Phys. Rev. E {\bf 88}, 033305 (2013)] where a multiscale numerical scheme based on kinetic model was developed for low speed isothermal flows with arbitrary Knudsen numbers. In this work, a discrete unified gas-kinetic scheme (DUGKS) for compressible flows with the consideration of heat transfer and shock discontinuity is developed based on the Shakhov model with an adjustable Prandtl number.
The method is an explicit finite-volume scheme where the transport and collision processes are coupled in the evaluation of the fluxes at cell interfaces,
so that the nice asymptotic preserving (AP) property is retained, such that the time step is limited only by the CFL number, the distribution function at cell interface recovers to the Chapman-Enskog one in the continuum limit while reduces to that of free-transport for free-molecular flow, and the time and spatial accuracy is of second-order accuracy in smooth region. These features make the DUGKS an ideal method for multiscale compressible flow simulations.
A number of numerical tests, including the shock structure problem, the Sod tube problem with different degree of non-equilibrium,
and the two-dimensional Riemann problem in continuum and rarefied regimes,
are performed to validate the scheme.
The comparisons with the results of DSMC and other benchmark data demonstrate that the DUGKS is a reliable and efficient method for multiscale compressible flow computation.
\end{abstract}

\maketitle
\section{Introduction}
It is a challenging problem for modeling and simulating non-equilibrium gas flows over a wide range of flow regimes.
The difficulty arises from the different temporal and spatial scales associated with the flows at different regimes.
For instance, for transition or free-molecule flows the typical time and length scales are the mean-collision time and the mean-free-path, respectively,
while for continuum flows the hydrodynamic scale is much larger than the kinetic scale.
For multiscale flows that involve different flow regimes, one popular numerical strategy is to use the hybrid approach,
which divides the flow domain into some subdomains and simulating the flow in each subdomain using different modeling according to the specific dynamics \cite{ref:Hadji2012}.
For instance, in hybrid particle-continuum approaches, the domain is divided into some Macro and Micro subdomains,
where particle-based methods such as molecular-dynamics (MD) or direct simulation
Monte Carlo (DSMC) are used in the Micro parts, while the continuum Navier-Stokes equations are used in the Macro parts.
Usually a buffer zone is employed in hybrid methods between neighboring subdomains to exchange flow information using different strategies \cite{ref:Thompson,ref:Werder,ref:Bjorn,ref:Mattew13,ref:Henry}.
A common feature of the hybrid method is that they are based on {\em numerical coupling} of solutions from different flow regimes,
which are limited to systems with a clear scale separation and may encounter significant difficulties for flows with a continuous scale variation \cite{ref:Hadji2012}.

Recently, some efforts have been made to develop numerical schemes for multiscale flows based on kinetic models (e.g. the Boltzmann equation or simplified models). Such kinetic schemes attempt to provide a unified description of flows in different regimes by discretizing the same kinetic equation dynamically,
so that the difficulties of hybrid methods in simulating cross-scale flows can be avoided. An example of kinetic schemes is the well-known discrete ordinate method (DOM) \cite{ref:DOM_Huang,ref:DOM_Li,ref:DOM_Kud}), which is powerful for flows in the kinetic regime, but may encounter difficulties for near continuum flow computation due to the limitation of small time step and large numerical dissipations.
In order to overcome this problem, some asymptotic preserving (AP) schemes were developed (e.g., \cite{ref:AP_Pie,ref:AP_Ben,ref:AP_Jin,ref:AP_Gia}),
which can recover the Euler solutions in the continuum regime, but may encounter difficulties for the Navier-Stokes solutions.
Therefore, it is still desirable to design kinetic schemes that can work efficiently for flows in a wide ranges of regimes.

The recent unified gas-kinetic scheme (UGKS) provides a dynamical multiscale method
which can get accurate solutions in both continuum and free molecular regimes \cite{ref:UGKS,ref:UGKS-Huang}.
The UGKS is a finite-volume scheme for the Boltzmann-BGK equation \cite{ref:BGK},
and the particle velocity is discretized into a discrete velocity set, like the DOM.
However, the update of the discrete distribution function considers the coupling of particle transport and collision
process in one time step, and so the time step is not limited by the collision term.
Furthermore, the UGKS adopts the local integral solution of the BGK equation in the reconstruction of the distribution function at cell interfaces for flux evaluation, which allows the scheme to change dynamically from the kinetic to hydrodynamic physics according to the local flow condition.
It is noted, however, in the original UGKS an additional evolutionary step for macroscopic variables is required such that extra computation costs are demanded.

An alternative simpler UGKS method, i.e., the so-called discrete unified gas kinetic scheme (DUGKS), was proposed recently \cite{ref:DUGKS}.
This scheme is also a finite-volume discretization of the Boltzmann-BGK equation. But unlike the UGKS \cite{ref:UGKS}, the evolution is based on a modified distribution function instead of the original one, which removes the implicitness in the update process of UGKS. At the same time, the evolution of macroscopic variables is not required any more. Furthermore, the distribution function at a cell interface is constructed based on the evolution equation itself instead of the local integral solution, so that the reconstruction is much simplified without scarifying the multiscale dynamics. The DUGKS has the same modeling mechanism as the original UGKS. The DUGKS has been applied successfully to a number of gas flows ranging from continuum to transition regimes \cite{ref:DUGKS}.

The previous DUGKS is designed for low-speed isothermal flows where the temperature variation is neglected.
In many non-equilibrium flows, however, temperature may change significantly (e.g. high Mach number flows)
or change differently from continuum flows (e.g. micro-flows). Under such circumstance,
it is necessary to track the temperature evolution in addition to the fluid dynamics.
In this work, a full DUGKS is developed for non-equilibrium gas flows where temperature variation is included.
The scheme is constructed based on the BGK-Shakhov model which can yield a correct Prandtl number in the continuum regime \cite{ref:Shakhov}.
The rest of this paper is organized as follows. In Sec. II, the full DUGKS is presented and some discussions on its properties are presented.
In Sec. III, a number of numerical tests, ranging from subsonic and hypersonic flows with different Knudsen numbers, are conducted to validate the method.
In Sec. IV, a brief summary is given.

\section{BGK-Shakhov model}
In kinetic theory, the BGK model \cite{ref:BGK} uses only one single relaxation time,
which leads to a fixed unit Prandtl number. In order to overcome this limitation,
a number of improved models, such as the BGK-Shakhov model \cite{ref:Shakhov} and the Ellipsoidal Statistical model
\cite{ref:ES}, have been proposed based on different physical consideration.
In $D$-dimensional space, the BGK-Shahkov model can be expressed as
\begin{equation}
\label{eq:Boltzmann-BGK}
\dpart{f}{t}+\bm{\xi}\cdot\nabla
f=\Omega\equiv -\dfrac{1}{\tau}\left[f-f^{S}\right],
\end{equation}
where $f=f(\x, \bm{\xi}, \bm{\eta}, \bm{\zeta}, t)$ is the velocity distribution function
for particles moving in $D$-dimensional physical space with velocity $\bm{\xi}=(\xi_1,\cdots,\xi_{\small{D}})$ at position $\x=(x_1, \cdots, x_{\small{D}})$ and
time $t$. Here $\bm{\eta}=(\xi_{{\small{D}}+1},\cdots,\xi_3)$ is a vector of length $L=3-D$, consisting of the rest components of the particle velocity  $(\xi_1,\xi_2,\xi_3)$ in 3-dimensional (3D) space; $\bm{\zeta}$ is a vector of length $K$ representing the internal degree of freedom of molecules; $\tau$ is the relaxation time relating to the dynamic viscosity $\mu$ and pressure $p$ with $\tau=\mu/p$, and $f^{S}$ is the Shakhov equilibrium distribution function given by
\begin{equation}
f^{S}=f^{eq}\left[1+(1-\mbox{Pr}) \dfrac{\c\cdot\q}{5pRT}\left(\dfrac{c^2+\eta^2}{RT}-5\right)\right]=f^{eq}+f_{\tPr},
\end{equation}
where $f^{eq}$ is the Maxwellian distribution function, $\mbox{Pr}$ is the Prandtl number, $\c=\bxi-\u$ is the peculiar velocity with $\u$ being the macroscopic flow velocity, $\q$ is the heat flux, $R$ is the gas constant, and $T$ is the temperature. The Maxwellian distribution function $f^{eq}$ is given by
\begin{equation}
f^{eq}=\dfrac{\rho}{(2\pi RT)^{(3+K)/2}}\exp\left(-\dfrac{c^2+\eta^2+\zeta^2}{2 RT }\right),
\end{equation}
where $\rho$ is the density. The conservative flow variables are defined by the moments of the distribution function,
\begin{equation}
\label{eq:W}
\bm{W}=\left(
\begin{array}{c}
\rho\\
\rho\bm{u}\\
\rho E
\end{array}
\right)= \int{\bm{\psi}(\bm{\xi}, \bm{\eta}, \bm{\zeta}) f\, d\bm{\xi}d\bm{\eta}d\bm{\zeta}},
\end{equation}
where $\bm{\psi}=(1, \bm{\xi}, (\xi^2+\eta^2+\zeta^2)/2)^T$ is the collision invariant,
 $\rho E=\rho u^2/2+\rho\epsilon$ is the total energy, and $\epsilon=c_{\small{V}} T$ is  the international energy with $c_{\small{V}}$ being the specific heat capacity at constant volume.
 The pressure is related to density and temperature through an ideal equation of state, $p=\rho R T$, and the heat flux is defined by
\begin{equation}
\q=\dfrac{1}{2}\int{\c (c^2+\eta^2+\zeta^2) f\,  d\bm{\xi}d\bm{\eta}d\bm{\zeta}}.
\end{equation}
The specific  heat capacities at constant pressure and volume are $c_{\small{p}}=(5+K)R/2$ and $c_{\small{V}}=(3+K)R/2$, respectively, and so the specific heat ratio is
\begin{equation}
\gamma=\dfrac{c_{\small{p}}}{c_{\small{V}}}=\dfrac{K+5}{K+3}.
\end{equation}
The stress tensor $\bm{\tau}$ is defined from the second-order moment of the distribution function,
\begin{equation}
\label{eq:tau-f}
\bm{\tau}=\int{\c \c f\,  d\bm{\xi}d\bm{\eta}d\bm{\zeta}}.
\end{equation}

The dynamic viscosity $\mu$ usually depends on the inter-molecular interactions. For example, for hard-sphere (HS) or variable hard-sphere
(VHS) molecules,
\begin{equation}
\mu=\mu_{ref}\left(\dfrac{T}{T_{ref}}\right)^{\omega},
\end{equation}
where $\omega$ is the index related to the HS or VHS model, $\mu_{ref}$ is the viscosity at the reference temperature $T_{ref}$.

The evolution of the distribution function $f$ depends only on the $D$-dimensional particle velocity $\bm{\xi}$ and is irrelevant to $\bm{\eta}$ and $\bm{\zeta}$.
In order to remove the dependence of the passive variables,  two reduced distribution functions can be introduced \cite{ref:DOM_Huang},
\begin{subequations}
\begin{equation}
g(\x,\bm{\xi},t)=\int{f(\x,\bm{\xi},\bm{\eta},\bm{\zeta}, t)d\bm{\eta}d\bm{\zeta}},
\end{equation}
\begin{equation}
h(\x,\bm{\xi},t)=\int{(\eta^2+\zeta^2)f(\x,\bm{\xi},\bm{\eta},\bm{\zeta}, t)d\bm{\eta}d\bm{\zeta}}.
\end{equation}
\end{subequations}
From Eq. \eqref{eq:W}, we can obtain that
\begin{equation}
\rho=\int{gd\bm{\xi}},\quad \rho\u=\int{\bm{\xi}gd\bm{\xi}},\quad \rho E=\dfrac{1}{2}\int{(\xi^2 g+h)d\bm{\xi}},
\end{equation}
and the heat flux $\bm{q}$ and the stress tensor can be computed as
\begin{equation}
\bm{q}=\dfrac{1}{2}\int{\bm{c}(c^2 g+h)d\bm{\xi}}, \quad\bm{\tau}=\int{\c \c g\,  d\bm{\xi}}.
\end{equation}

The evolution equations for $g$ and $h$ can be obtained from Eq. \eqref{eq:Boltzmann-BGK},
\begin{subequations}
\label{eq:BGK-gh}
\begin{equation}
\label{eq:BGK-g}
\dpart{g}{t}+\bm{\xi}\cdot\nabla g=\Omega_g\equiv -\dfrac{1}{\tau}\left[g-g^{S}\right],
\end{equation}
\begin{equation}
\label{eq:BGK-h}
\dpart{h}{t}+\bm{\xi}\cdot\nabla h=\Omega_h\equiv -\dfrac{1}{\tau}\left[h-h^{S}\right],
\end{equation}
\end{subequations}
where the reduced Shakhov distribution functions $g^S$ and $h^S$ are given by
\begin{subequations}
\label{eq:reduced-BGK}
\begin{equation}
g^S(\x,\bm{\xi},t)=\int{f^S(\x,\bm{\xi},\bm{\eta},\bm{\zeta}, t)d\bm{\eta}d\bm{\zeta}}=g^{eq}+g_{\tPr},
\end{equation}
\begin{equation}
h^S(\x,\bm{\xi},t)=\int{(\eta^2+\zeta^2)f^S(\x,\bm{\xi},\bm{\eta},\bm{\zeta}, t)d\bm{\eta}d\bm{\zeta}}=h^{eq}+h_{\tPr} ,
\end{equation}
\end{subequations}
with
\begin{subequations}
\begin{equation}
g^{eq}=\int{f^{eq}d\bm{\eta}d\bm{\zeta}}=\dfrac{\rho}{(2\pi RT)^{D/2}}\exp\left[-\dfrac{(\bxi-\u)^2}{2RT}\right],
\end{equation}
\begin{equation}
h^{eq}=\int{(\eta^2+\zeta^2)f^{eq}d\bm{\eta}d\bm{\zeta}}=(K+3-D)RT g^{eq},
\end{equation}
\begin{equation}
g_{\tPr}=\int{f_{\tPr}d\bm{\eta}d\bm{\zeta}}=(1-\mbox{Pr})\dfrac{\c\cdot\q}{5pRT}\left[\dfrac{c^2}{RT}-D-2\right]g^{eq},
\end{equation}
and
\begin{equation}
h_{\tPr}=\int{(\eta^2+\zeta^2)f_{\tPr}d\bm{\eta}d\bm{\zeta}}=(1-\mbox{Pr})\dfrac{\c\cdot\q}{5pRT}\left[\left(\dfrac{c^2}{RT}-D\right)(K+3-D)-2K\right]RT g^{eq}.
\end{equation}
\end{subequations}
With the definitions of the conserved variables, it is easy to verify that the collision terms $\Omega_g$ and $\Omega_h$ satisfy the following conservative laws,
\begin{equation}
\label{eq:Conserve-Omega}
\int{\Omega_g\d\bxi}=0,\quad \int{\bxi\Omega_g\d\bxi}=\bm{0},\quad \int{(\xi^2\Omega_g+\Omega_h)\d\bxi}=0.
\end{equation}

\section{Discrete unified gas kinetic scheme}
\subsection{Updating of the cell-averaged distribution function}
The full DUGKS is constructed based on the two reduced kinetic equations \eqref{eq:reduced-BGK}. The scheme is a finite volume formulation of the kinetic equations. For simplicity, we rewrite Eq. \eqref{eq:BGK-gh} in the following form,
\begin{equation}
\label{eq:BGK-phi}
\dpart{\phi}{t}+\bm{\xi}\cdot\nabla \phi=\Omega\equiv -\dfrac{1}{\tau}\left[\phi-\phi^{S}\right],
\end{equation}
for $\phi=g$ or $h$. The domain is decomposed into a set of control volumes (cells),
then the integration of Eq. \eqref{eq:BGK-phi} over cell $j$ centering at $\x_j$ from time $t_n$ to $t_{n+1}$  with time step $\Delta t$ leads to
\begin{equation}
\label{eq:cell_phi}
 \phi_j^{n+1}(\bxi)-\phi_j^n(\bxi) + \dfrac{\Delta t}{|V_j|}{\bm{F}}^{n+1/2}(\bxi)=\dfrac{\Delta t}{2}\left[\Omega_j^{n+1}(\bxi)+\Omega_j^n(\bxi)\right],
\end{equation}
where the midpoint rule is used for the time integration of the convection
term, and the trapezoidal rule for the collision term. Such treatment ensures the scheme is of second-order accuracy in time. Here
\begin{equation}
\label{eq:Flux}
 {\bm{ F}}^{n+1/2}(\bxi)=\int_{\partial V_j
}{(\bm{\xi}\cdot\bm{n}) \phi(\bm{x}, \bxi, t_{n+1/2})\, d\bm{S}}
\end{equation}
is the micro flux across the cell interface, where $|V_j|$ and $\partial
V_j$ are the volume and surface of cell $V_j$, $\bm{n}$ is the
outward unit vector normal to the surface, and $\phi_j$ and
$\Omega_j$ are the cell-averaged values of the distribution
function and collision term, respectively, e.g.,
$$
\phi_j^n(\bxi)=\dfrac{1}{|V_j|}\int_{V_j}{\phi(\x, \bxi, t_n) d\x}.
$$

The update rule given by Eq. \eqref{eq:cell_phi} is implicit due to the term $\Omega_j^{n+1}$ which requires the conserved variables $\bm{W}_j^{n+1}$. In order to remove this implicity, we employ a technique as used in the development of the isothermal DUGKS \cite{ref:DUGKS}, i.e., we introduce a new distribution function,
\begin{equation}
\label{eq:tilde-phi}
\tilde{\phi}=\phi-\dfrac{\Delta t}{2}\Omega=\dfrac{2\tau+\Delta t
}{2\tau} \phi-\dfrac{\Delta t}{2\tau}\phi^{S}.
\end{equation}
Then Eq. \eqref{eq:cell_phi} can be rewritten as
\begin{equation}
\label{eq:GKS-tilde-phi}
 \tilde{\phi}_j^{n+1}=\tilde{\phi}_j^{+,n} - \dfrac{\Delta t}{|V_j|}{\bm{F}}^{n+1/2},
\end{equation}
where
\begin{equation}
\label{eq:tilde-phi+}
\tilde{\phi}^+=\phi+\dfrac{\Delta t}{2}\Omega=\dfrac{2\tau-\Delta t}{2\tau+\Delta
t}\tilde{\phi}+\dfrac{2\Delta t}{2\tau+\Delta t}\phi^{S}.
\end{equation}

It is noted that from the conservative properties of the collision operators given by Eq. \eqref{eq:Conserve-Omega}, we can obtain that
\begin{equation}
\label{eq:W-tilde}
\rho=\int{\tilde{g} d\bm{\xi}},\quad \rho\u=\int{\bm{\xi}\tilde{g}d\bm{\xi}},\quad \rho E=\dfrac{1}{2}\int{(\xi^2 \tilde{g}+\tilde{h})d\bm{\xi}}.
\end{equation}
Therefore, in practical computations we can track the distribution function $\tilde{g}$ and $\tilde{h}$ instead of the original ones, which can evolve explicitly according to Eq. \eqref{eq:GKS-tilde-phi}, provided the micro-flux $\bm{F}$ at the cell interface at $t_{n+1/2}$ is obtained. In addition to the conserved variables, the heat flux $\q$ and stress tensor $\bm{\tau}$ can also be obtained from $\tilde{\phi}$. Actually, it can be shown that
\begin{equation}
\label{eq:q-tilde}
\q=\dfrac{2\tau}{2\tau+\Delta t \mbox{Pr}} \tilde{\q}, \quad \mbox{with} \quad \tilde{\q}=\dfrac{1}{2}\int{\bm{c}(c^2 \tilde{g}+\tilde{h})d\bm{\xi}}.
\end{equation}
\begin{equation}
\label{eq:tau-tilde}
\bm{\tau}=\dfrac{2\tau}{2\tau+\Delta t} \tilde{\bm{\tau}}, \quad \mbox{with} \quad \tilde{\bm{\tau}}=\int{\c\c \tilde{g}d\bm{\xi}}.
\end{equation}

\subsection{Flux evaluation}
The key in evaluating $\bm{F}^{n+1/2}$ is to reconstruct the distribution function $f^{n+1/2}$ at the cell interface. To do so we integrate Eq. \eqref{eq:BGK-phi} along the characteristic line within a half time step  $s=\Delta t/2$,
\begin{equation}
\label{eq:phi-face0}
 \phi\left(\x_b,\bm{\xi},t_{n}+s\right)-\phi\left(\x_b-\bm{\xi}s,\bm{\xi},t_{n}\right) =\dfrac{s
}{2}\left[\Omega\left(\x_b,\bm{\xi},t_{n}+s\right)+\Omega\left(\x_b-\bm{\xi}s,\bm{\xi},t_{n}\right)\right],
\end{equation}
where $\x_b\in\partial V_j$ is a point at the interface of cell $j$, and the trapezoidal rule is again used to evaluate the collision
term. It is noted that the formulation \eqref{eq:phi-face0} is also implicit due to the collision term $\Omega_j^{n+1}$. Similar to the treatment for $\tilde{\phi}$, we introduce another distribution function $\bar{\phi}$ to remove the implicity,
\begin{equation}
\label{eq:bar-phi}
 \bar{\phi}=\phi-\dfrac{s}{2}\Omega=\dfrac{2\tau+s
}{2\tau} \phi -\dfrac{s}{2\tau}\phi^{S}.
\end{equation}
Then Eq. (\ref{eq:phi-face0}) can be rewritten as
\begin{equation}
\label{eq:phi-bar}
 \bar{\phi}\left(\x_b,\bm{\xi},t_{n+1/2}\right)=\bar{\phi}^+(\x_b-\bm{\xi}s,\bm{\xi},t_{n}),
\end{equation}
where
\begin{equation}
\label{eq:phi-bar+}
\bar{\phi}^+=\phi+\dfrac{s}{2}\Omega=\dfrac{2\tau-s}{2\tau+s}\bar{\phi}+\dfrac{2s}{2\tau+s}\phi^{S}.
\end{equation}
Therefore, once $\bar{\phi}^+(\x_b-\bm{\xi}s,\bm{\xi},t_{n})$ is obtained, the distribution function $\bar{\phi}(\x_b, \bxi, t_{n+1/2})$ can be determined from Eq. \eqref{eq:phi-bar}. It is noted that the conserved variables can also be obtained from $\bar{g}$ and $\bar{h}$ like Eq. \eqref{eq:W-tilde},
\begin{equation}
\label{eq:W-bar}
\rho=\int{\bar{g} d\bm{\xi}},\quad \rho\u=\int{\bm{\xi}\bar{g}d\bm{\xi}},\quad \rho E=\dfrac{1}{2}\int{(\xi^2 \bar{g}+\bar{h})d\bm{\xi}},
\end{equation}
which means that $\bm{W}(\x_b,t_{n+1/2})$ can be obtained from $\bar{\phi}(\x_b,\bxi,t_{n+1/2})$ directly. Furthermore, the heat flux $\q(\x_b,t_{n+1/2})$ can also be determined from  $\bar{\phi}(\x_b,\bxi,t_{n+1/2})$,
\begin{equation}
\label{eq:q-bar}
\q=\dfrac{2\tau}{2\tau+s \mbox{Pr}} \bar{\q}, \quad\mbox{with}\quad \bar{\q}=\dfrac{1}{2}\int{\bm{c}(c^2 \bar{g}+\bar{h})d\bm{\xi}}.
\end{equation}
Then the Shakhov distribution $\phi^S$ at cell interface $\x_b$ and time $t_{n+1/2}$ can be evaluated, and subsequently the original distribution function can be calculated from Eq. \eqref{eq:bar-phi} as
\begin{equation}
\label{eq:phi-face}
\phi(\x_b, \bxi, t_{n+1/2})=\dfrac{2\tau} {2\tau+s}\bar{\phi}(\x_b,\bxi, t_{n}+s)+\dfrac{s}{2\tau+s }\phi^{S}(\x_b,\bxi, t_n+s).
\end{equation}

\begin{figure}
\includegraphics[width=0.45\textwidth]{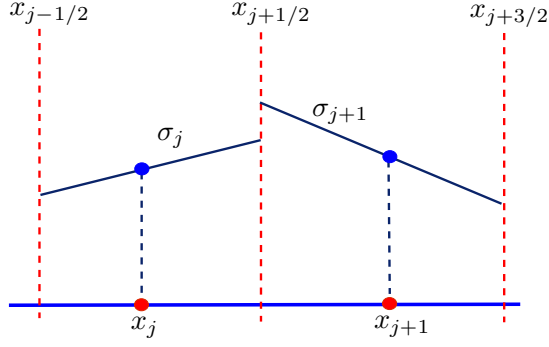}
\caption{(Color online) Schematic of 1D cell geometry.} \label{fig:Cell}
\end{figure}

Now the task is to determine the $\bar{\phi}^+(\x_b-\bm{\xi}s,\bm{\xi},t_{n})$. This is achieved through a reconstruction of the profile of $\bar{\phi}^+(t_n)$ in each cell. First, we determine the the cell-averaged distribution function $\bar{\phi}^+(t_n)$ at the cell center $\x_j$ from the tracked distribution function $\tilde{\phi}(\x_j, t_n)$. From Eqs. \eqref{eq:tilde-phi},\eqref{eq:bar-phi}, and \eqref{eq:phi-bar+}, we can obtain that
\begin{equation}
\label{eq:A_bar-phi+}
 \bar{\phi}^+=\dfrac{2\tau-s}{2\tau+\Delta
t}\tilde{\phi}+ \dfrac{3s}{2\tau+\Delta t}\phi^{S}.
\end{equation}
It should be noted that $\tilde{\phi}^+$ and  $\bar{\phi}^+$ are related. Actually, from Eqs. \eqref{eq:tilde-phi+} and \eqref{eq:A_bar-phi+} we can obtain that
\begin{equation}
\label{eq:A_tilde-phi+}
\tilde{\phi}^+=\dfrac{4}{3}\bar{\phi}^+-\dfrac{1}{3}\tilde{\phi}.
\end{equation}
With this relation the computation can be simplified as noted in the following subsection.

Assuming that in each cell $\bar{\phi}^+$ is linear, then we have
\begin{equation}
\label{eq:bar-phi+}
\bar{\phi}^+(\x_b-\bm{\xi}s,\bm{\xi},t_{n})=
\bar{\phi}^+(\x_j,\bxi, t_n)+(\x_b-\x_j-\bm{\xi}s)\cdot\bm{\sigma}_j,\quad (\x_b-\bm{\xi}s) \in V_j,
\end{equation}
where $\bm{\sigma}_j$ is the slope of $\bar{\phi}^+$ in cell $j$. As an example, in Fig. \ref{fig:Cell} a 1D case is shown. In this case, in order to reconstruct the distribution function $\phi$ at the cell interface $x_b=x_{j+1/2}$, the distribution function $\bar{\phi}^+$ is approximated as
\begin{equation}
\bar{\phi}^+(x_b-\xi s,\xi, t_n)=
\begin{cases}
\bar{\phi}^+(x_j,\xi, t_n)+(x_b-\xi s -x_j)\sigma_j, & \xi >0, \\
\bar{\phi}^+(x_{j+1},\xi, t_n)+(x_b-\xi s -x_{j+1})\sigma_{j+1}, & \xi <0.
\end{cases}
\end{equation}
The slope $\bm{\sigma}_j$ in each cell can be reconstructed from the cell-averaged values using some numerical limiters. For example, in the 1D case shown in Fig. \ref{fig:Cell}, we can use the van Leer limiter \cite{ref:vanLeer}, i.e.,
\begin{equation}
\label{eq:slope}
\sigma_j=\left[\mbox{sign}(s_1)+\mbox{sign}(s_2)\right]\dfrac{|s_1||s_2|}{|s_1|+|s_2|},
\end{equation}
where
\begin{equation}
s_1=\dfrac{\bar{\phi}_j^+ - \bar{\phi}_{j-1}^+}{x_j-x_{j-1}},\quad s_2=\dfrac{\bar{\phi}_{j+1}^+ - \bar{\phi}_{j}^+}{x_{j+1}-x_{j}}.
\end{equation}

\subsection{Evolution procedure}
In summary, the procedure of the DUGKS at each time step $t_n$ can be listed as follows (assuming $\x_b$ is the cell interface of cell $j$ centered at $\x_j$):
\renewcommand{\labelenumi}{(\arabic{enumi})}
\renewcommand{\labelenumii}{(\alph{enumii})}
\begin{enumerate}
\item Calculate the micro flux $\bm{F}$ at cell interface $\x_b$ and at time $t_{n+1/2}$
\begin{enumerate}
  \item Calculate $\bar{\phi}^+$ from $\tilde{\phi}$ at each cell center with velocity $\bxi$ according to Eq. \eqref{eq:A_bar-phi+};
  \item Reconstruct the gradient of $\bar{\phi}^+$ (i.e., $\bm{\sigma}$) in each cell using certain numerical limiters, e.g., Eq. \eqref{eq:slope} in 1D case;
  \item Reconstruct the distribution function $\bar{\phi}^+$ at $\x_b-\bxi s$ according to Eq. \eqref{eq:bar-phi+};
  \item Determine the distribution function $\bar{\phi}$ at cell interface at time $t_{n+1/2}$ according to Eq. \eqref{eq:phi-bar};
  \item Calculate the conserved variables $\bm{W}(\x_b,t_{n+1/2})$ and heat flux $\bm{q}(\x_b,t_{n+1/2})$ from $\bar{\phi}(\x_b,\bxi, t_{n+1/2})$, see Eqs. \eqref{eq:W-bar} and \eqref{eq:q-bar};
  \item Calculate the original distribution function $\phi$ at cell interface and $t_{n+1/2}$ from $\bar{\phi}(\x_b,\bxi, t_{n+1/2})$ and $\phi^S(\x_b,\bxi, t_{n+1/2})$ according to Eq. \eqref{eq:phi-face};
  \item Calculate the micro flux $\bm{F}^{n+1/2}$ through each cell interface from $\phi^{n+1/2}$ according to Eq. \eqref{eq:Flux};
  \end{enumerate}
  \item Calculate $\tilde{\phi}^+$ at cell center and time $t_n$ according to Eq. \eqref{eq:A_tilde-phi+};
  \item Update the cell-averaged $\tilde{\phi}$ in each cell from $t_n$ to $t_{n+1}$ according to Eq. \eqref{eq:cell_phi}.
\end{enumerate}

The particle velocity $\bxi$ is continuous in the above procedure. In practical computations, the velocity space will be discretized into a set of discrete velocities $\bxi_i$ $(i=1, 2, \cdots, b)$. Usually the discrete velocity set is chosen as the abscissas of certain quadrature rules such as the Gaussian-Hermite or Newton-Cotes formula, and the integrals in the above procedure will be replaced by the quadrature. For example, the conserved variables can be computed as
\begin{equation}
\rho=\sum_{i=1}^b{w_i\tilde{g}(\bxi_i)}, \quad \rho\u=\sum_{i=1}^b{w_i\bxi_i\tilde{g}(\bxi_i)}, \quad \rho E=\dfrac{1}{2}\sum_{i=1}^b{w_i\left[\xi^2\tilde{g}(\bxi_i)+\tilde{h}(\bxi)\right]},
\end{equation}
where $w_i$ is the associate quadrature weights.

\section{Analysis of DUGKS}
We now discuss some important properties of the DUGKS. First, we will show the DUGKS has the asymptotic preserving (AP) property \cite{ref:UGKS,ref:MieussensRAD}, namely (i) the time step $\Delta t$ is independent of the particle collision time for all Knudsen numbers, and (ii) the scheme is consistent with the Navier-Stokes equations in the continuum limit. Regarding the time step, it is noted that the particle transport and collisions are coupled in the reconstruction of the interface distribution function, which is necessary for an AP scheme \cite{ref:UGKS}.
This coupling also releases the constraint on the collision-time and the time step as in the operator-splitting schemes,
and the time-step can be determined by the Courant-Friedrichs-Lewy (CFL) condition \cite{ref:UGKS,ref:DUGKS},
\begin{equation}
\Delta t = \alpha \dfrac{\Delta x}{U_m+\xi_m},
\end{equation}
where $\alpha$ is the CFL number, $\Delta x$ is the minimal grid
spacing, $\xi_m$ is the maximum discrete velocity,and $U_m$ is the maximum flow velocity. $\Delta t$ determined in this way does not dependent on the relaxation time $\tau$, and the DUGKS is uniformly stable with respect to the Knudsen number.

Regarding point (ii), it is noted that in the continuum limit as $\tau\ll\Delta t$, the distribution function in a cell given by Eq. \eqref{eq:bar-phi+} can be approximated as
\begin{equation}
\label{eq:AP-bar-phi+}
\bar{\phi}^+(\x_b-\bm{\xi}s,\bm{\xi},t_{n})=
\bar{\phi}^+(\x_b,\bxi, t_n)-s\bm{\xi}\cdot\bm{\sigma}_{b}+O(\Delta x^2),
\end{equation}
where $\bm{\sigma}_{b}$ is the slope of $\bar{\phi}^+(\bxi,t_n)$ at the cell interface $\x_b$. Furthermore, follow the procedure given in the Appendix B of Ref. \cite{ref:DUGKS}, we can show that
\begin{subequations}
\begin{equation}
\phi(\x_b,\bxi, t)= \phi^S(\x_b,\bxi, t) -\tau D_t \phi^{S}(\x_b,\bxi, t) +O(\partial^2),
\end{equation}
\begin{equation}
\phi^S(\x_b,\bxi, t_n+s)= \phi^S(\x_b,\bxi, t_n) +s\partial_t \phi^{S}(\x_b,\bxi, t_n) +O(\partial^2).
\end{equation}
\end{subequations}
Then, with the aids of these results, we can obtain from Eqs. \eqref{eq:phi-bar}, \eqref{eq:phi-bar+}, and \eqref{eq:phi-face} that (refer to Appendix B of Ref. \cite{ref:DUGKS})
\begin{equation}
\phi(\x_b,\bm{\xi},t_n+s)\approx
\phi^{S}(\x_b,\bm{\xi},t_n)-\tau(\partial_t+\bxi\cdot\nabla)\phi^{S}(\x_b,\bm{\xi},t_n)+s\partial_t
\phi^{S}(\x_b,\bm{\xi},t_n),
\end{equation}
which recovers the Chapman-Enskog approximation for the Navier-Stokes solution  \cite{ref:GKS,ref:UGKS}. This fact suggests that the DUGKS can be viewed as a Navier-Stokes solver in the continuum limit.
It is also note that the use of the mid-point and trapezoidal rules in Eqs. \eqref{eq:cell_phi} and
\eqref{eq:phi-face0} as well as the linear reconstruction of the distribution function at the cell interface ensures a second-order accuracy in both space and time in the continuum limit.

On the other hand, in the free-molecule limit where $\tau\gg\Delta
t=2s$, we can find from Eq. \eqref{eq:phi-bar+} that $\bar{\phi}^+(\x_b-\bxi s,\bxi,t_n)\approx \bar{\phi}(\x_b-\bxi s,\bxi,t_n)$, and then from Eq. \eqref{eq:phi-bar} that $\bar{\phi}(\x_b,\bxi,t_n+s)=\bar{\phi}^+(\x_b-\bxi s,\bxi,t_n)\approx \bar{\phi}(\x_b-\bxi s,\bxi,t_n)$.
Furthermore, the relationship between $\bar{\phi}$ and $\phi$ as shown in Eq. \eqref{eq:bar-phi} gives that
$\phi(\x_b, \bxi, t_n+s)\approx\bar{\phi}(\x_b,\bxi,t_n+s)\approx f(x_b-\bm{\xi}s,t_n)$, which is just the
collision-less limit.

Finally we point out some key differences between the present DUGKS and the UGKS \cite{ref:UGKS,ref:UGKS-Huang} which is also designed for all Knudsen number flows, although both share many common features such as multi-dimensional nature, AP property, and coupling of particle transport and collision.
The first key difference is that the cell-averaged conserved variables $\bm{W}$ and heat flux $\bm{q}$ in each cell are required to evolve along with the cell-averaged distribution functions in the UGKS, because the collision term is discretized with the trapezoidal rule and the evaluation of the implicit part needs these quantities. However, with the newly introduced distribution function $\tilde{\phi}$, the implicity in the collision term is removed in the DUGKS, and $\bm{W}$ and $\bm{q}$ are not required to evolve. The second key difference between DUGKS and UGKS lies in the reconstruction of the distribution function at cell interfaces. In the UGKS \cite{ref:UGKS,ref:UGKS-Huang}, the interface distribution function $\phi(\x_b, t)$ is constructed based on the integral solution of the kinetic equation with certain approximations, while in the present DUGKS it is constructed based on the characteristic solution which is much simpler. The third difference is that the DUGKS is solely based on the single relaxation kinetic model due to its combination of the distribution function and the collision term, but the UGKS can be extended to the full Boltzmann collision term as well \cite{ref:Boltzmann}.
 Despite of these differences, we will show in next section that the present DUGKS can yield numerical predictions nearly the same as the UGKS.

\begin{figure}[h!]
\includegraphics[width=0.48\textwidth]{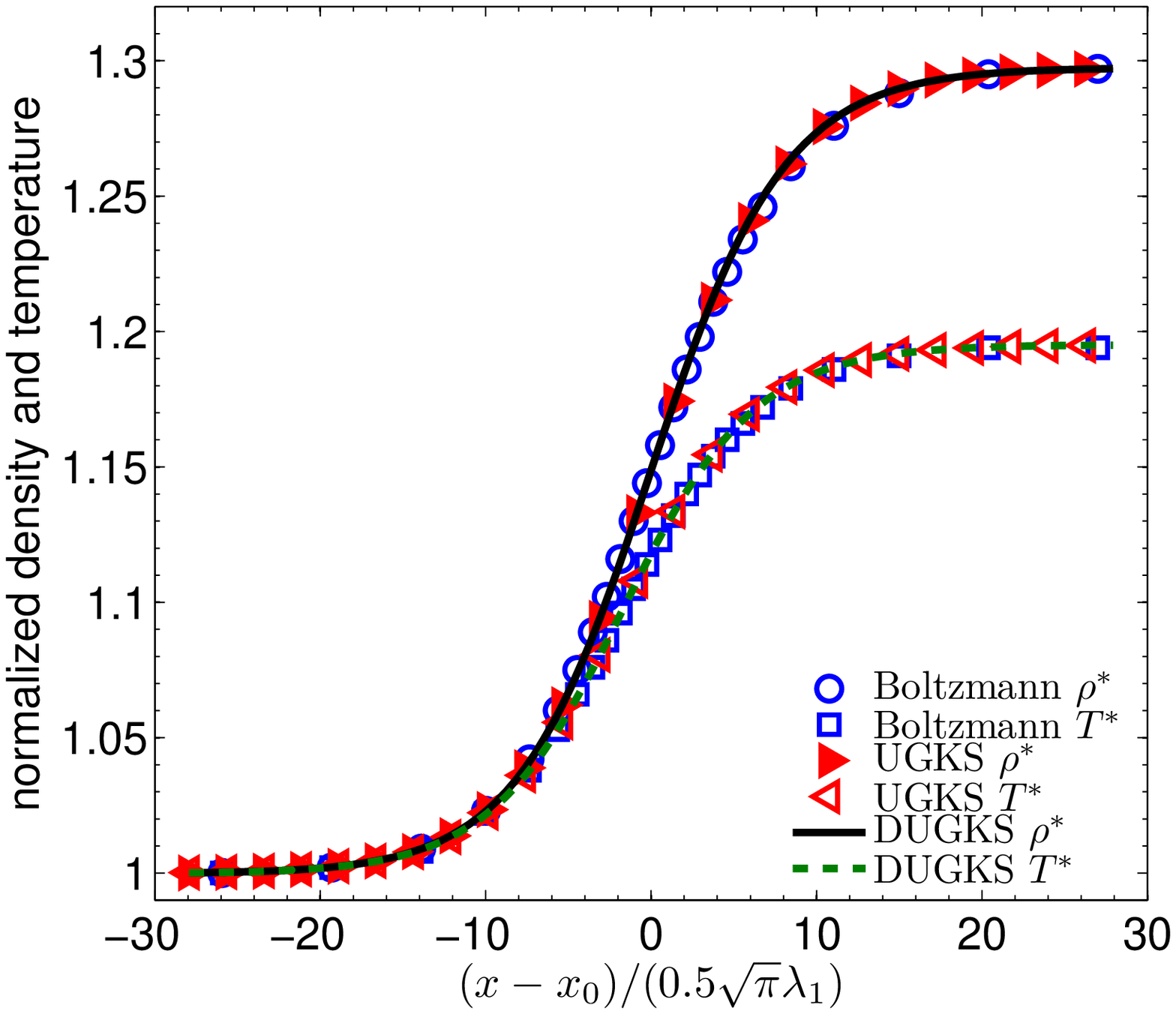}\hfill
\includegraphics[width=0.48\textwidth]{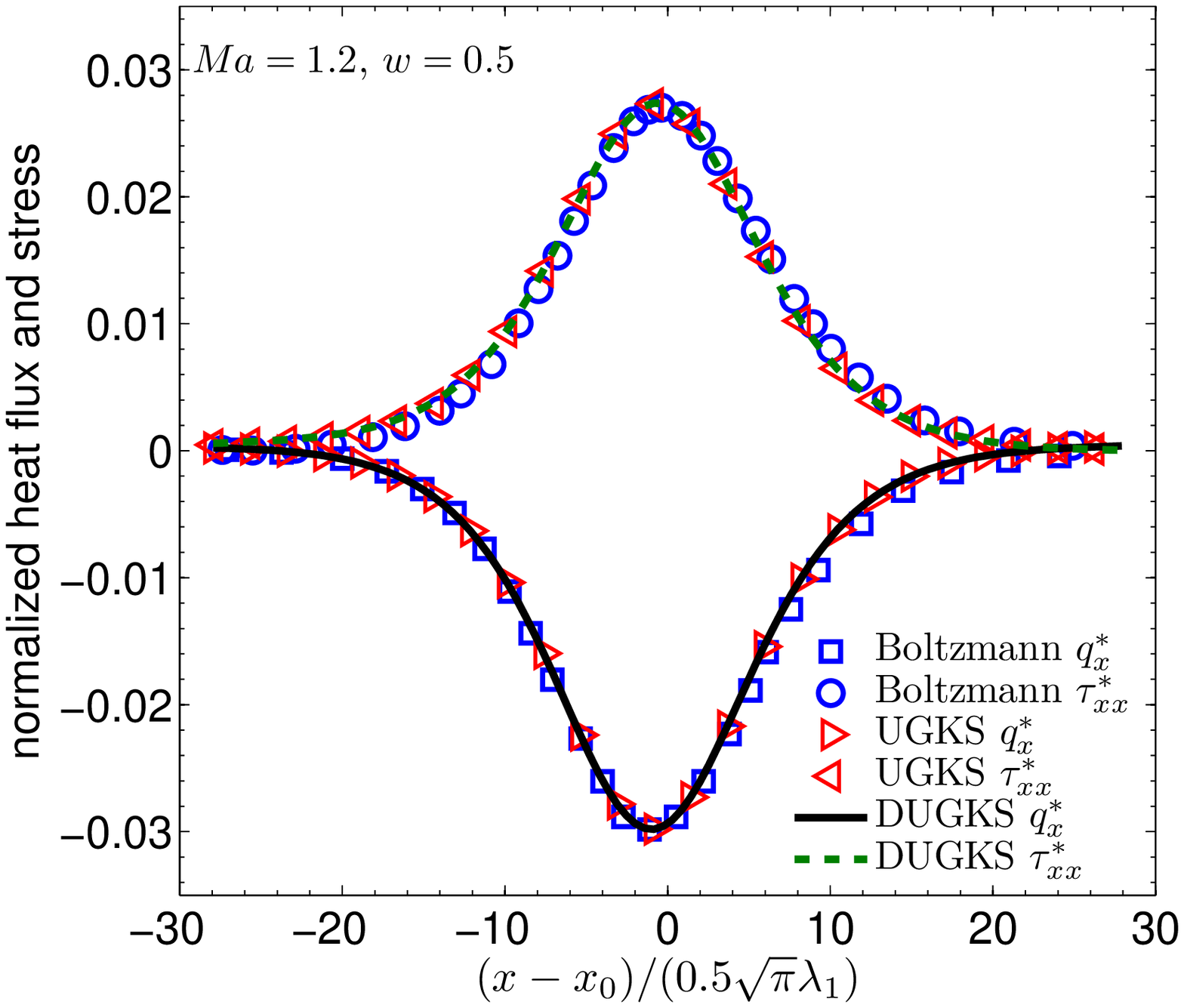}\vfill
\includegraphics[width=0.48\textwidth]{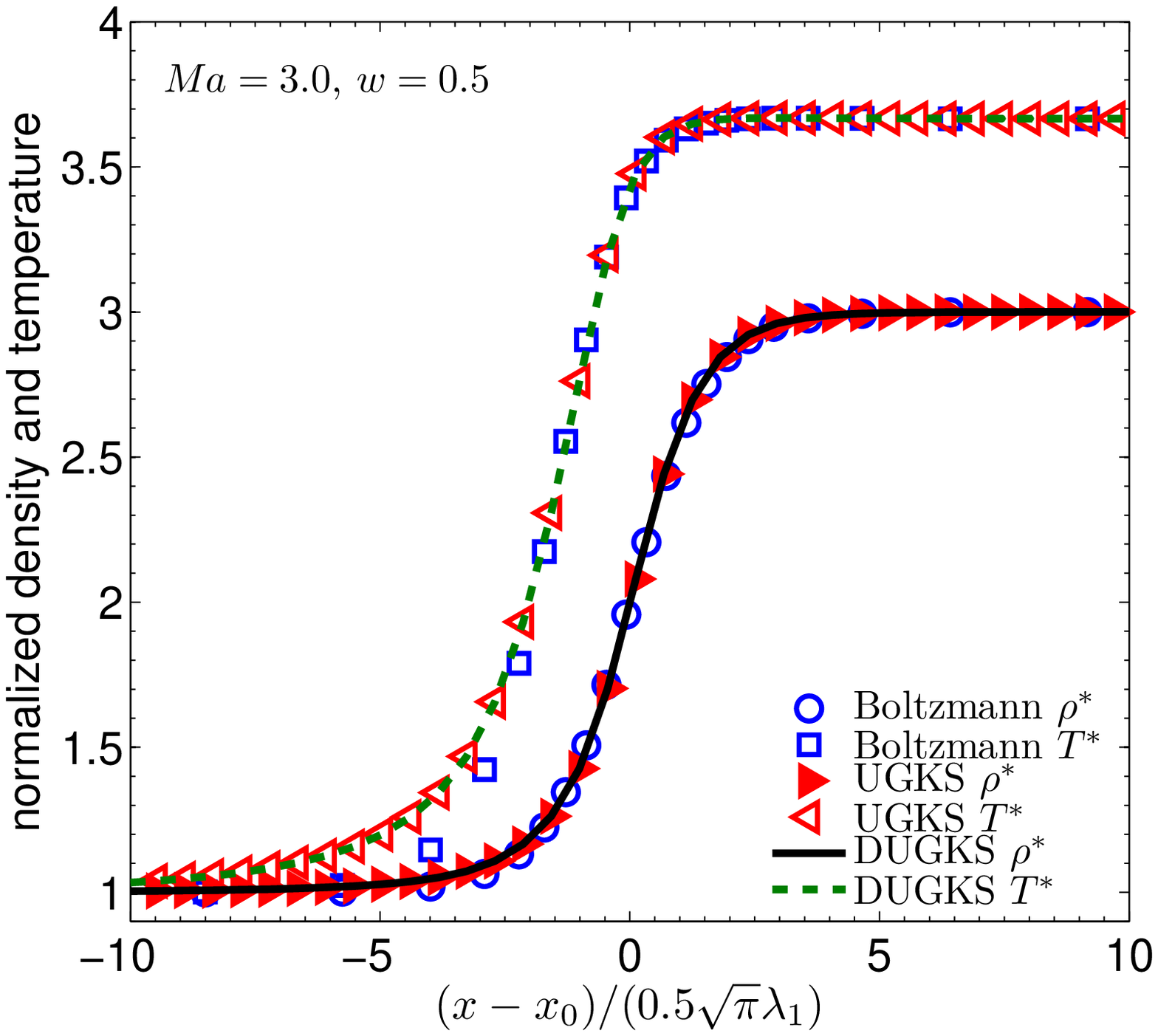}\hfill
\includegraphics[width=0.48\textwidth]{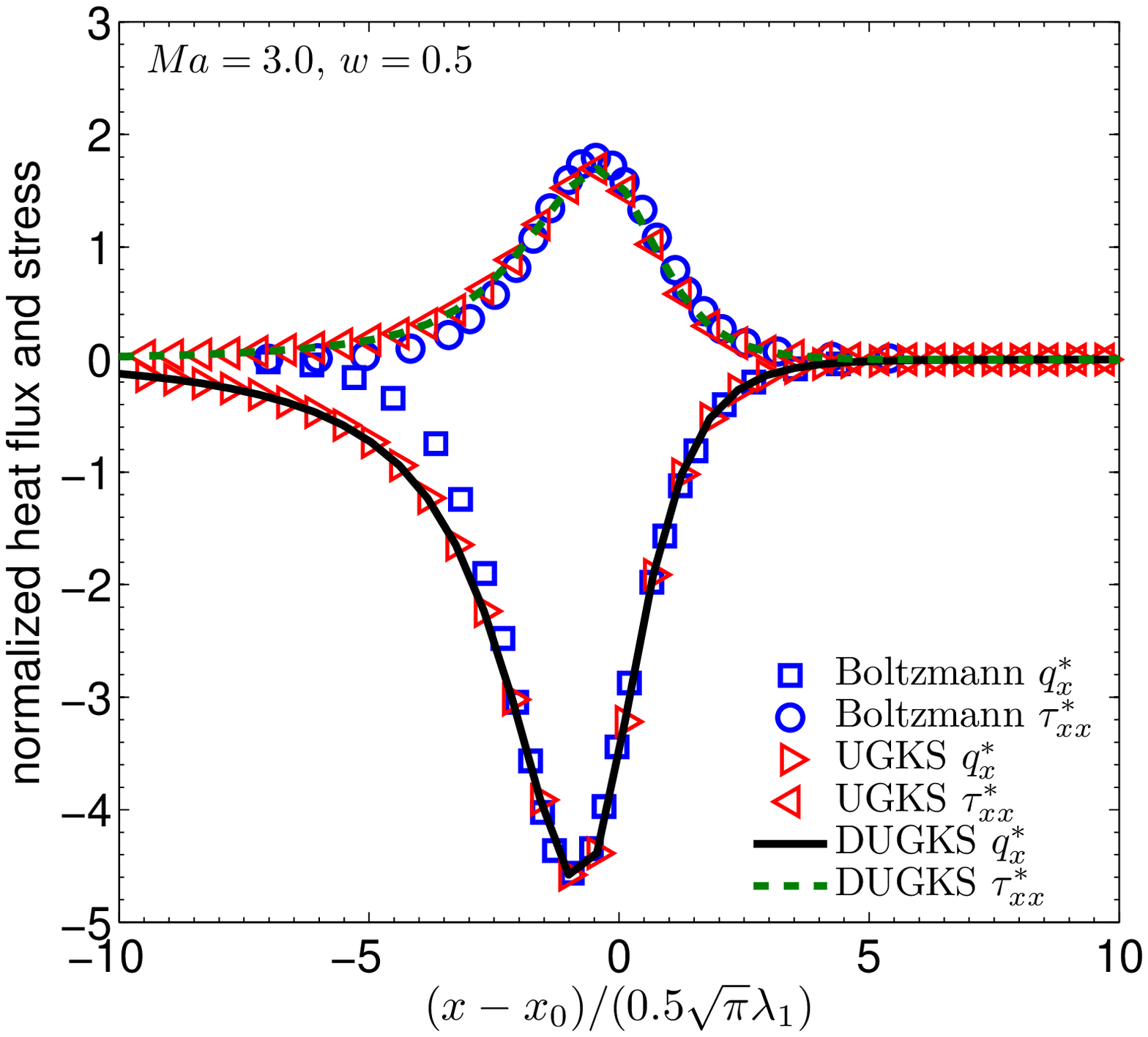}
\caption{(Color online) Shock structure of hard-sphere gas at different Mach numbers. Left: density ($\rho^*$) and temperature ($T^*$); Right: Stress ($\tau_{xx}^*$) and heat flux ($q_x^*$). } \label{fig:HS05}
\end{figure}

\begin{figure}[t!]
\includegraphics[width=0.48\textwidth]{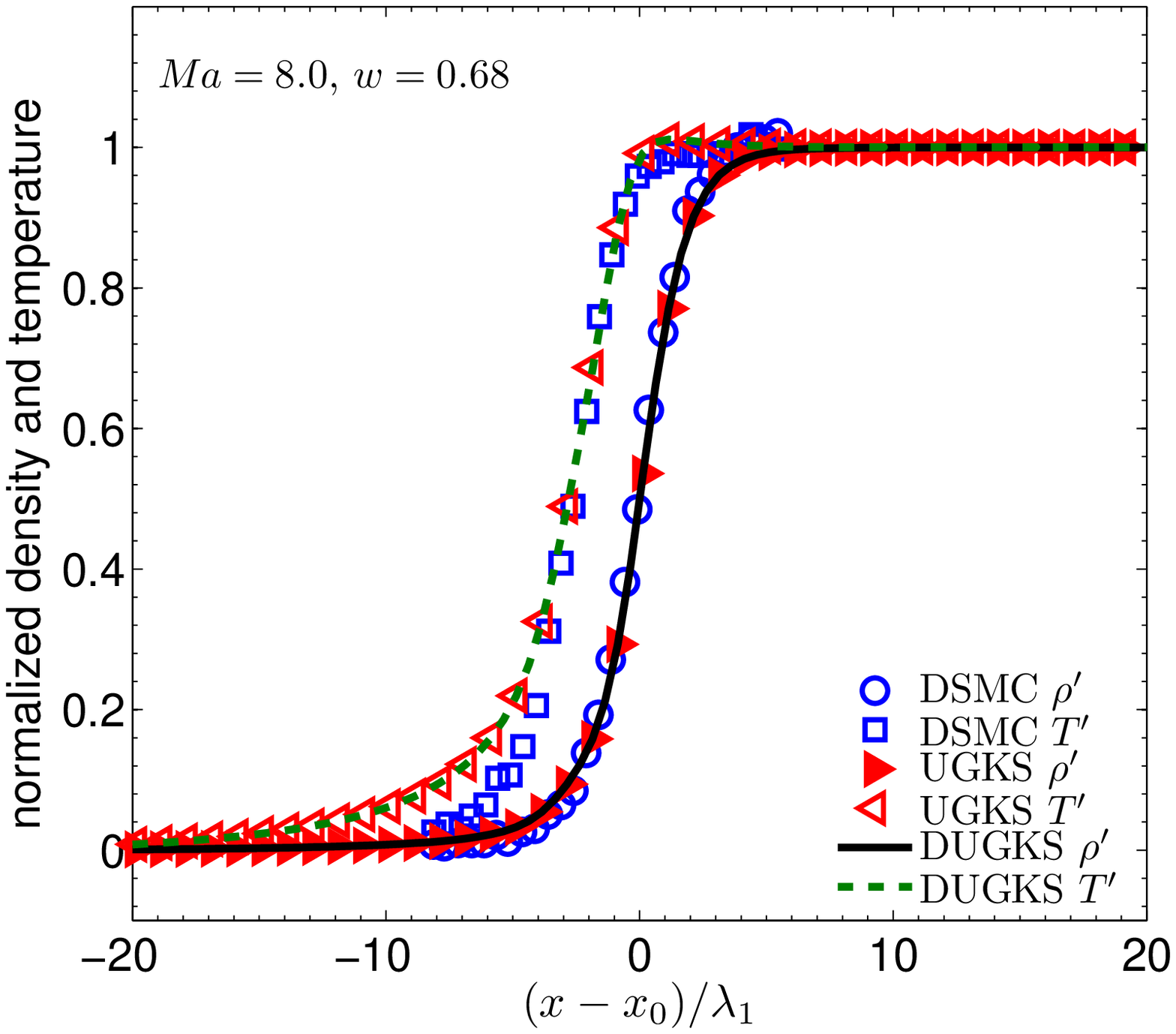}\hfill
\includegraphics[width=0.48\textwidth]{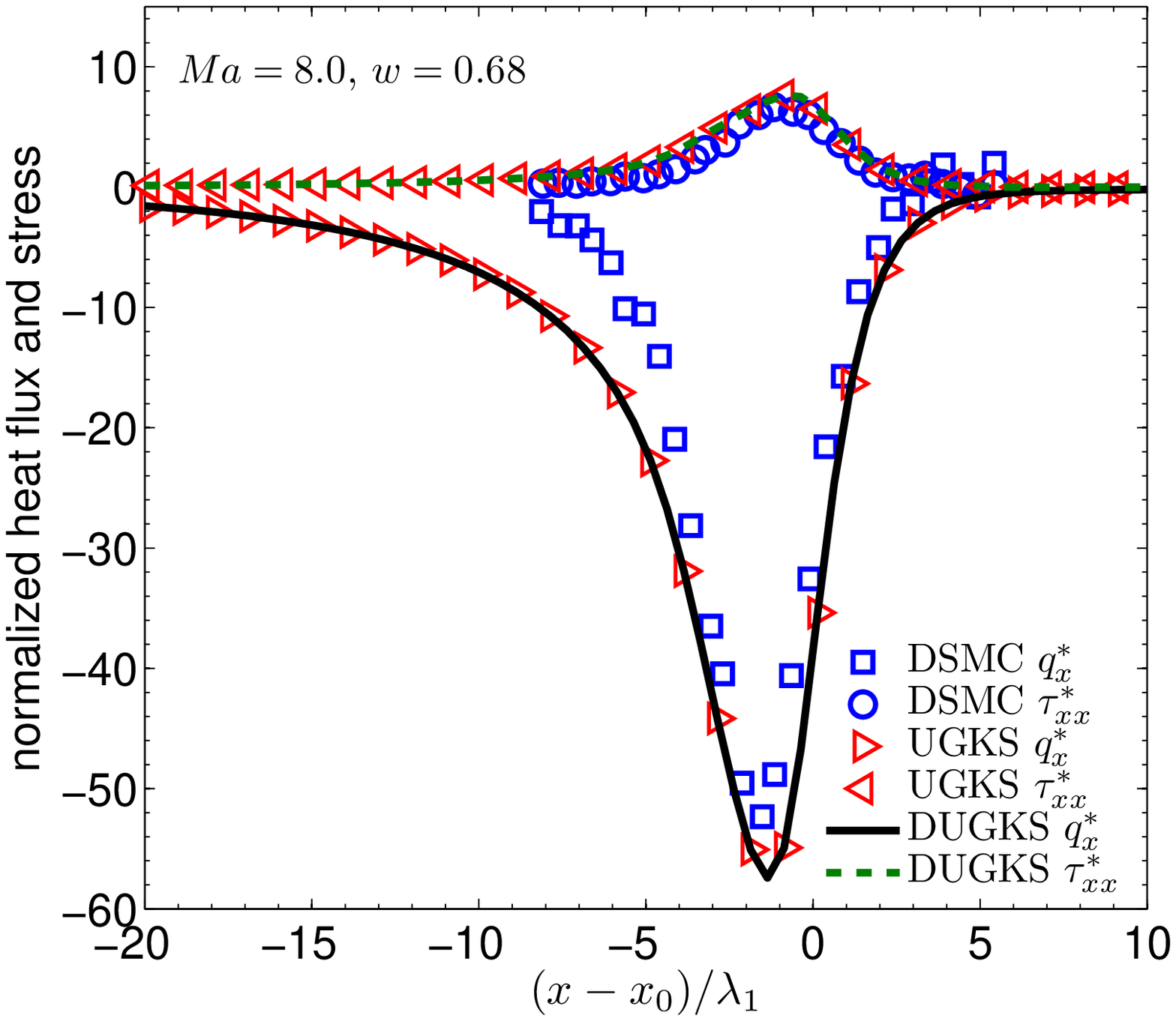}
\caption{(Color online) Shock structure with $\mbox{Ma}=8$ and $w=0.68$. Left: density ($\rho'$) and temperature ($T'$); Right: Stress ($\tau_{xx}^*$) and heat flux ($q_x^*$). } \label{fig:Ma8}
\end{figure}

\begin{figure}
\includegraphics[width=0.48\textwidth]{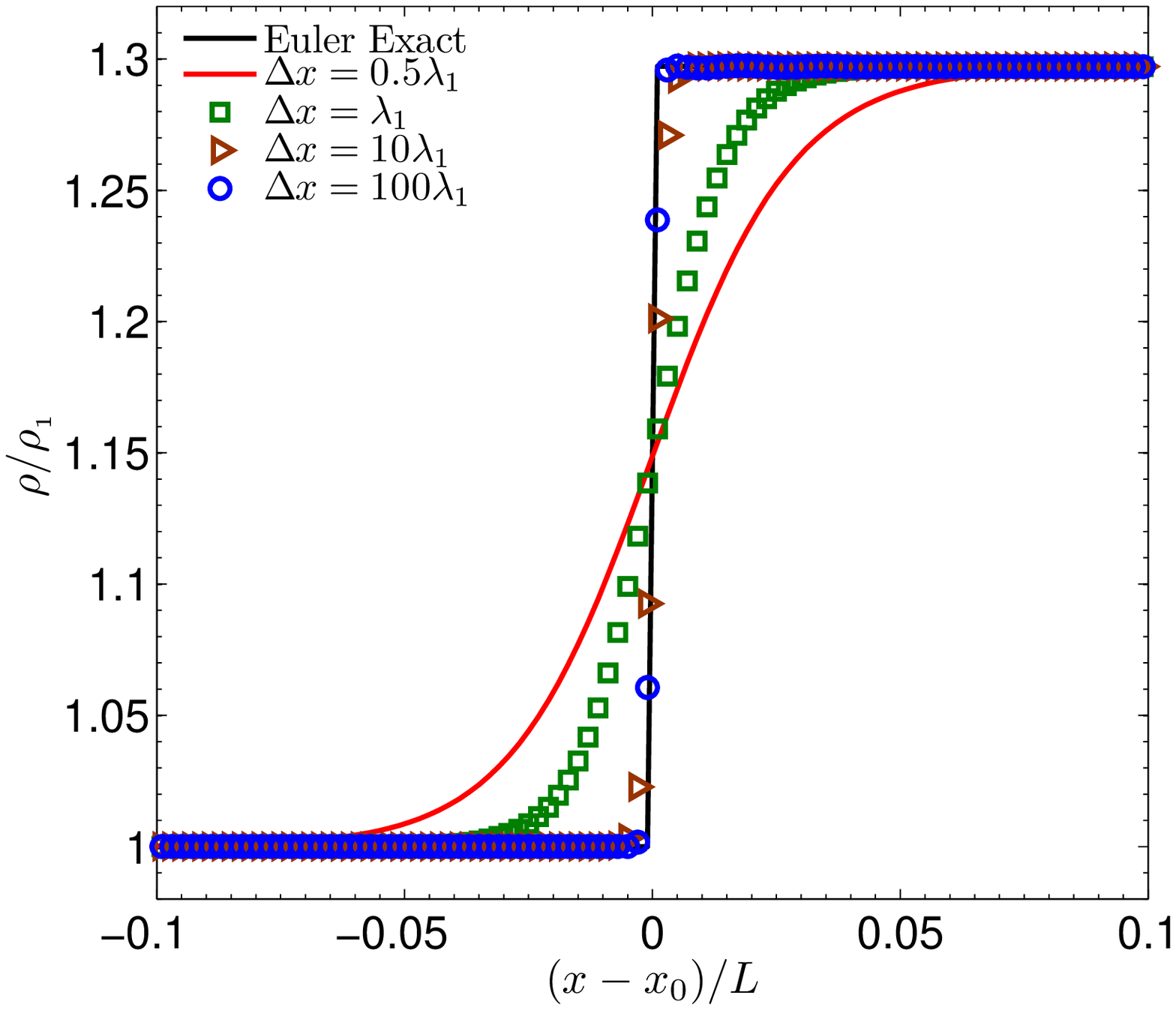}\hfill
\includegraphics[width=0.48\textwidth]{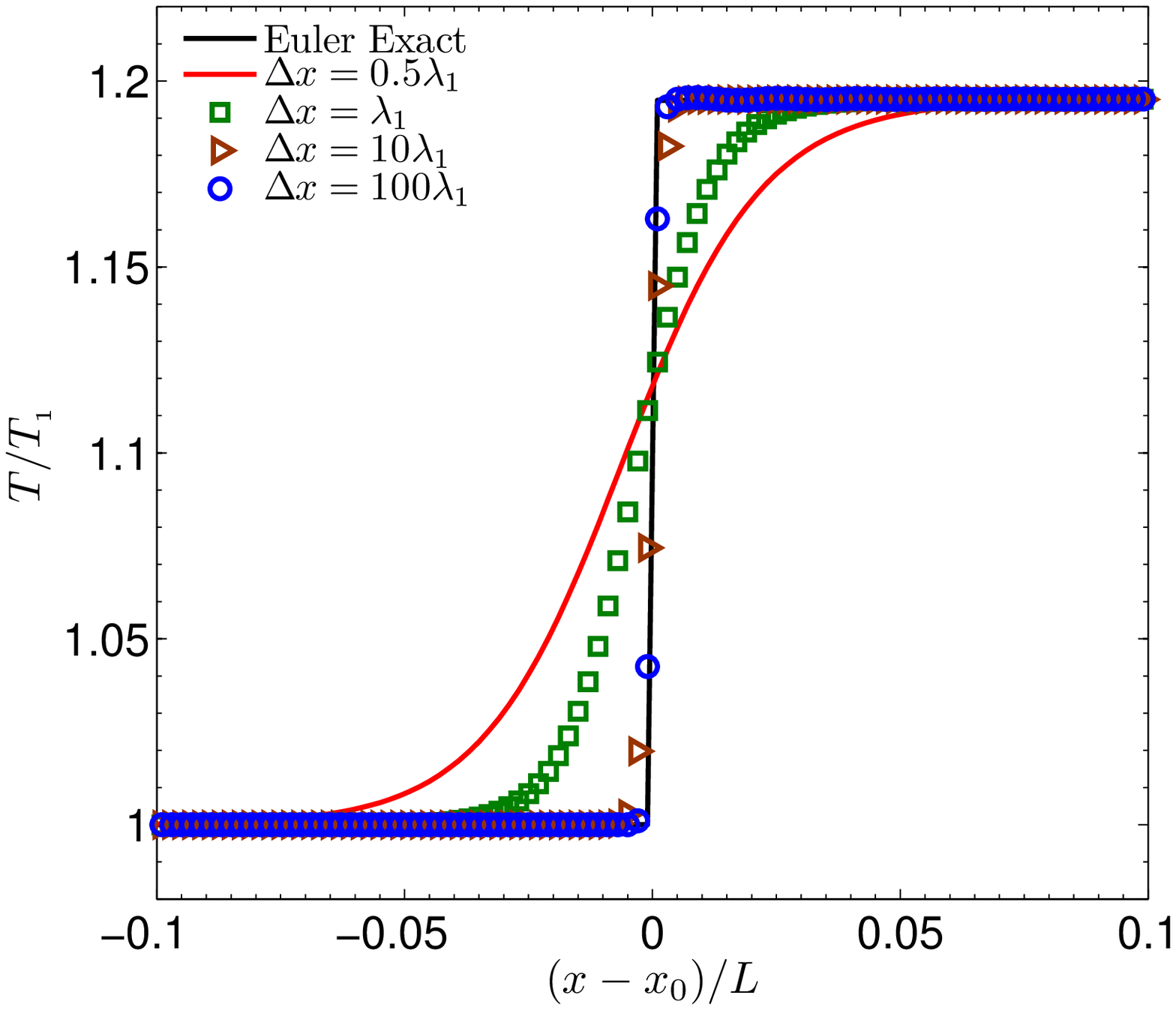}
\caption{(Color online) Density and Temperature profiles of the shock structure ($\mbox{Ma}=1.2$, $w=0.5$) with different cell sizes and CFL number 0.95.} \label{fig:MaCapture}
\end{figure}

\section{numerical tests}
The present DUGKS will be validated by a number of test problems in different flow regimes in this section. The problems include 1D and 2D subsonic/supersonic flows. In the simulations the van Leer limiter \cite{ref:vanLeer} will be used in the reconstruction of interface distribution function.

\subsection{1D shock structure}
The first test case is the argon shock structure from low to high Mach numbers.
The results of the present DUGKS simulations will be compared with the Boltzmann solution, DSMC result, and UGKS prediction.
The densities, velocities, and temperatures at upstream ($\rho_1$, $u_1$, $T_1$) and downstream ($\rho_2$, $u_2$, $T_2$) satisfy the Rankine-Hugoniou conditions \cite{ref:Harris}. The Prandtl number and specific heat ratio for argon are $\mbox{Pr}=2/3$ and $\gamma=5/3$, respectively, and the viscosity depends on the temperature, $\mu\varpropto T^w$, where $w$ relates to the inter-molecular interactions \cite{ref:Harris}. The mean-free-path $\lambda$ is related to the viscosity as \cite{ref:Bird},
\begin{equation}
\lambda=\dfrac{2\mu(7-2w)(5-2w)}{15\rho(2\pi RT)^{1/2}}.
\end{equation}

In the simulations the flow variables are normalized by the corresponding upstream quantities, and the characteristic density, length, velocity, and time are choosen to be $\rho_1$, $\lambda_1$, $\sqrt{2RT_1}$, and $\lambda_1/\sqrt{2RT_1}$, respectively. The computational domain is chosen to be $-25\lambda_1\le x \le 25\lambda_1$. A uniform mesh with 100 cells is used so that the mesh space is $\Delta x=0.5\lambda_1$. The discrete velocity set is determined by the Newton-Cotes quadrature with 101 points distributed uniformly in $[-15, 15]$. Initially, the distribution functions at $x\le 0$ are set to be the Maxwellian distribution with the upstream state, and those at $x>0$ are set to be the Maxwellian distribution with the downstream state. The CFL number used in all simulations is set to be 0.95.

First we consider hard-sphere model (i.e. $w=0.5$), which was also studied by Ohwada by solving the full Boltzmann equation numerically \cite{ref:Ohwada}, and by Xu and Huang using the UGKS method with the Shakhov model \cite{ref:UGKS_Shk}. In Fig. \ref{fig:HS05} the profiles of the normalized density $\rho^*=\rho/\rho_1$, temperature $T^*=T/T_1$, heat flux $q_x^*=q_x/p_1$, and shear stress $\tau_{xx}^*=\tau_{xx}/p_1(2RT_1)^{3/2}$ with $p_1=\rho_1 R T_1$, are shown at $\mbox{Ma}=1.2$ and 3.0. Here the the location of the shock is chosen to be $x_0$ such that $\rho(x_0)=(\rho_1+\rho_2)/2$, the heat flux and stress are computed according to Eqs. \eqref{eq:q-tilde} and \eqref{eq:tau-tilde}. The results are compared with those of the Boltzmann and UGKS solutions. It can be observed that the overall agreement between the present numerical results are in good agreement with the previous studies. Particularly, it is found that the difference between the present DUGKS and the previous UGKS \cite{ref:UGKS_Shk} are rather small. For instance, at $\mbox{Ma}=3.0$ the maximum relative differences in density and temperature between the two solutions are about $8.09\times 10^{-5}$ and $2.24\times 10^{-4}$, respectively.

We next test the shock structure of $\mbox{Ma}=8$ with $w=0.68$ as studied by the DSMC \cite{ref:Ma8-DSMC} and UGKS \cite{ref:UGKS_Shk} methods. In Fig. \ref{fig:Ma8} the normalized density $\rho'=(\rho-\rho_1)/(\rho_2-\rho_1)$, temperature $T'=(T-T_1)/(T_2-T_1)$, heat flux $q_x^*$, and
stress $\tau_{xx}^*$, are shown and compared with the DSMC and UGKS data \cite{ref:Ma8-DSMC,ref:UGKS_Shk}. Again the results predicted by the present DUGKS are in close agreement with those of the UGKS method, and both compare well with the DSMC results in general. However, it is noted that the temperature profiles predicted by the present DUGKS and UGKS methods increase earlier in the upstream than that of the DSMC, and accordingly the heat flux also decreases earlier (heat flux is in the negative direction). The shear stress $\tau_{xx}$ predicted by the present DUGKS agrees well with the DSMC result,
and the peak values of the heat flux predicted by the three methods are quite close.

The present DUGKS is also tested as a shock capturing scheme. This is achieved by varying the cell size $\Delta x$ with a fixed CFL number, as suggested in \cite{ref:UGKS}.
As an example, the $\mbox{Ma}=1.2$ shock structure of a hard-sphere gas with a fixed upstream mean-free-path ($\lambda_1=1.0$) is simulated by the present DUGKS.
In the calculations, the cell size $\Delta x$ changes from $0.5\lambda_1$ to $100\lambda_1$, and the CFL number is fixed at 0.95 so that the time step changes with the cell size accordingly.
In Fig. \ref{fig:MaCapture} the density and temperature profiles are shown with different cell sizes.
As observed, the solution goes from well-resolved to highly under-resolved solution with increasing of cell sizes.
Particularly, as $\Delta x=100\lambda_1$ the solution agrees very well with the exact solution of the Euler equations,
suggesting that the DUGKS becomes an effective shock capturing scheme in this case.
The low dissipative nature of the DUGKS is due to the coupling of collision and transport in the reconstruction of the cell interface flux.
It is noted that the discrete ordinate method (DOM) may encounter numerical instability with $\Delta x=100\lambda_1$ and a CFL number 0.95 \cite{ref:UGKS}, and a much smaller CFL number should be used to obtain a stable but more dissipative solution.
\begin{figure}
\includegraphics[width=0.48\textwidth]{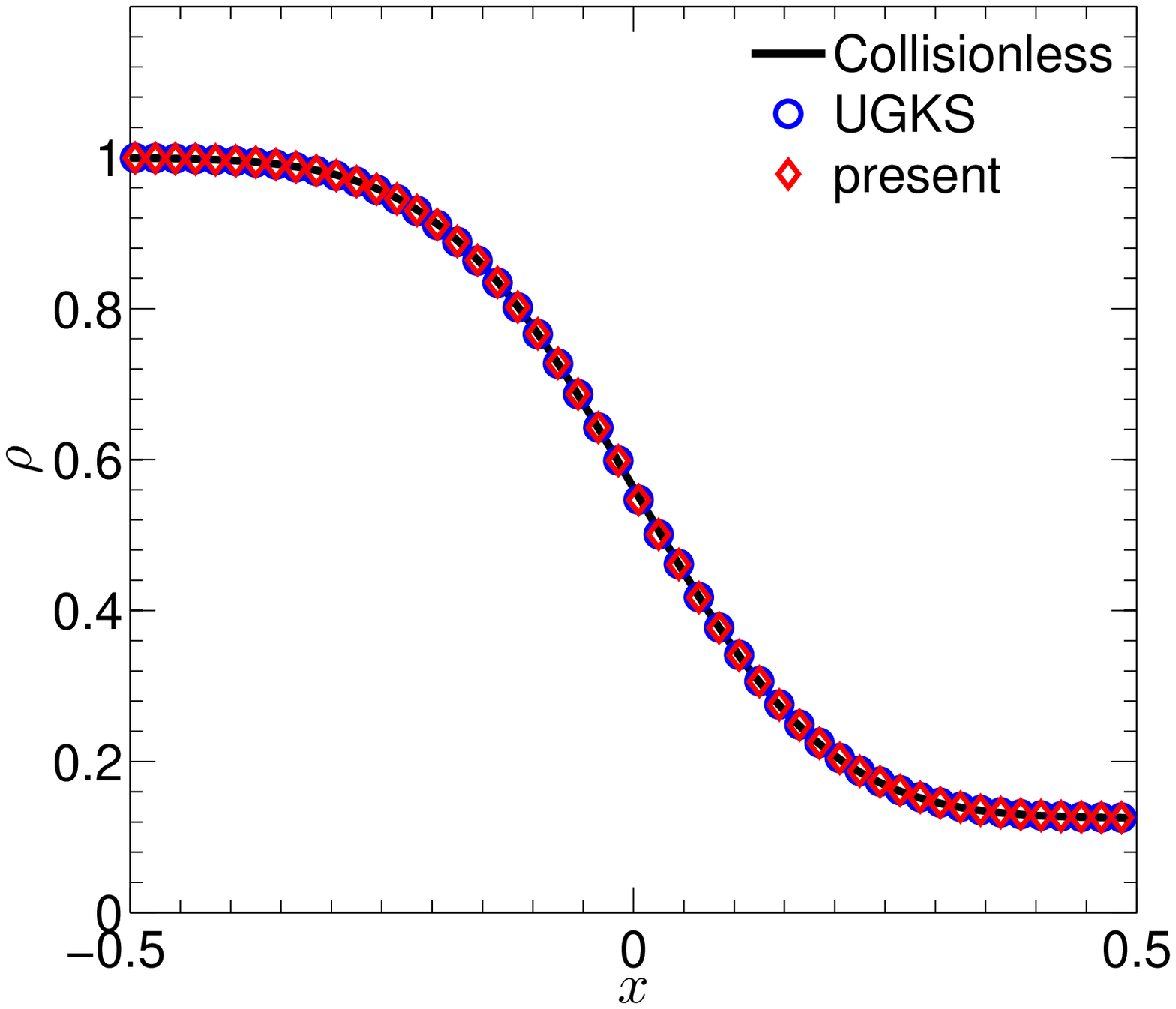}\vfill
\includegraphics[width=0.48\textwidth]{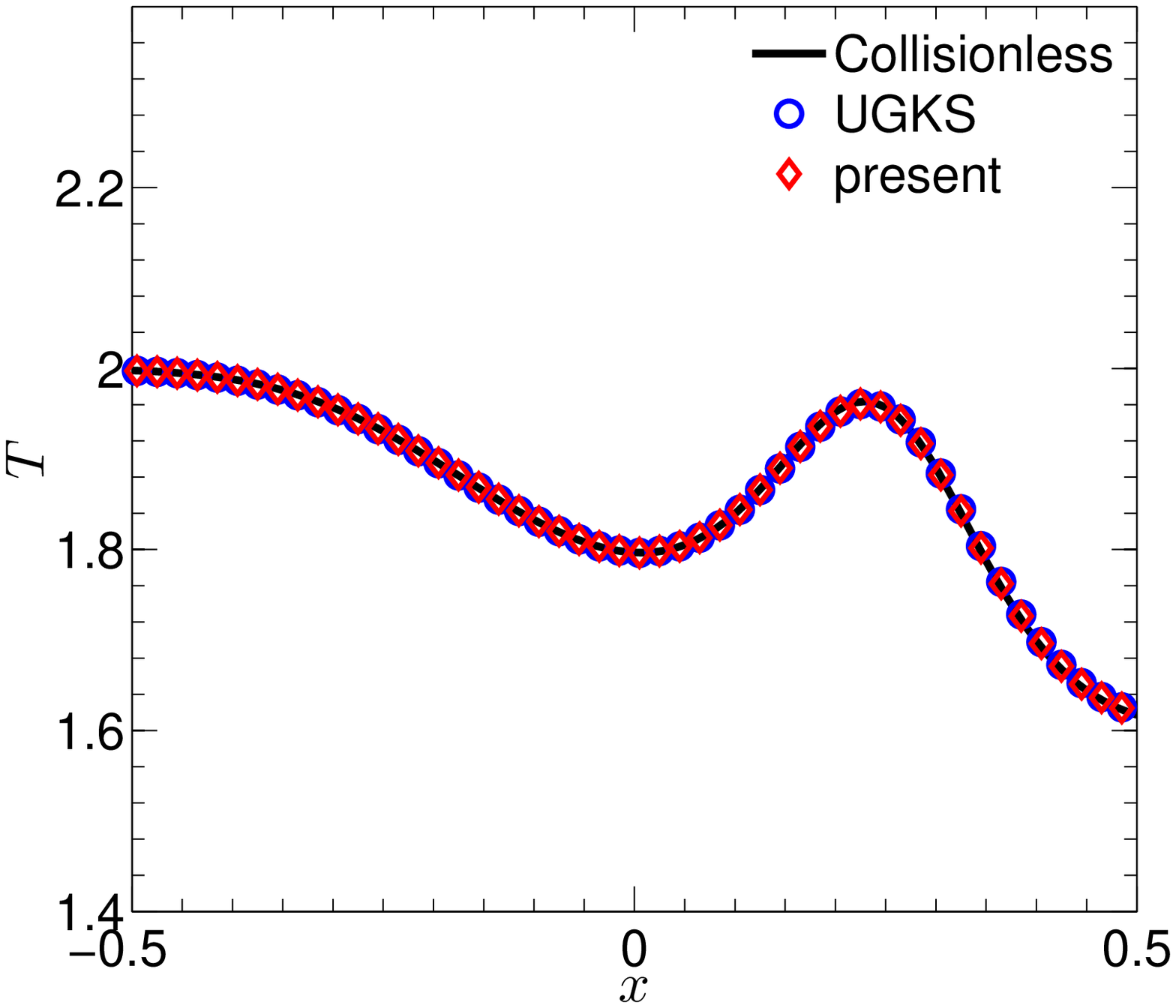}\vfill
\includegraphics[width=0.48\textwidth]{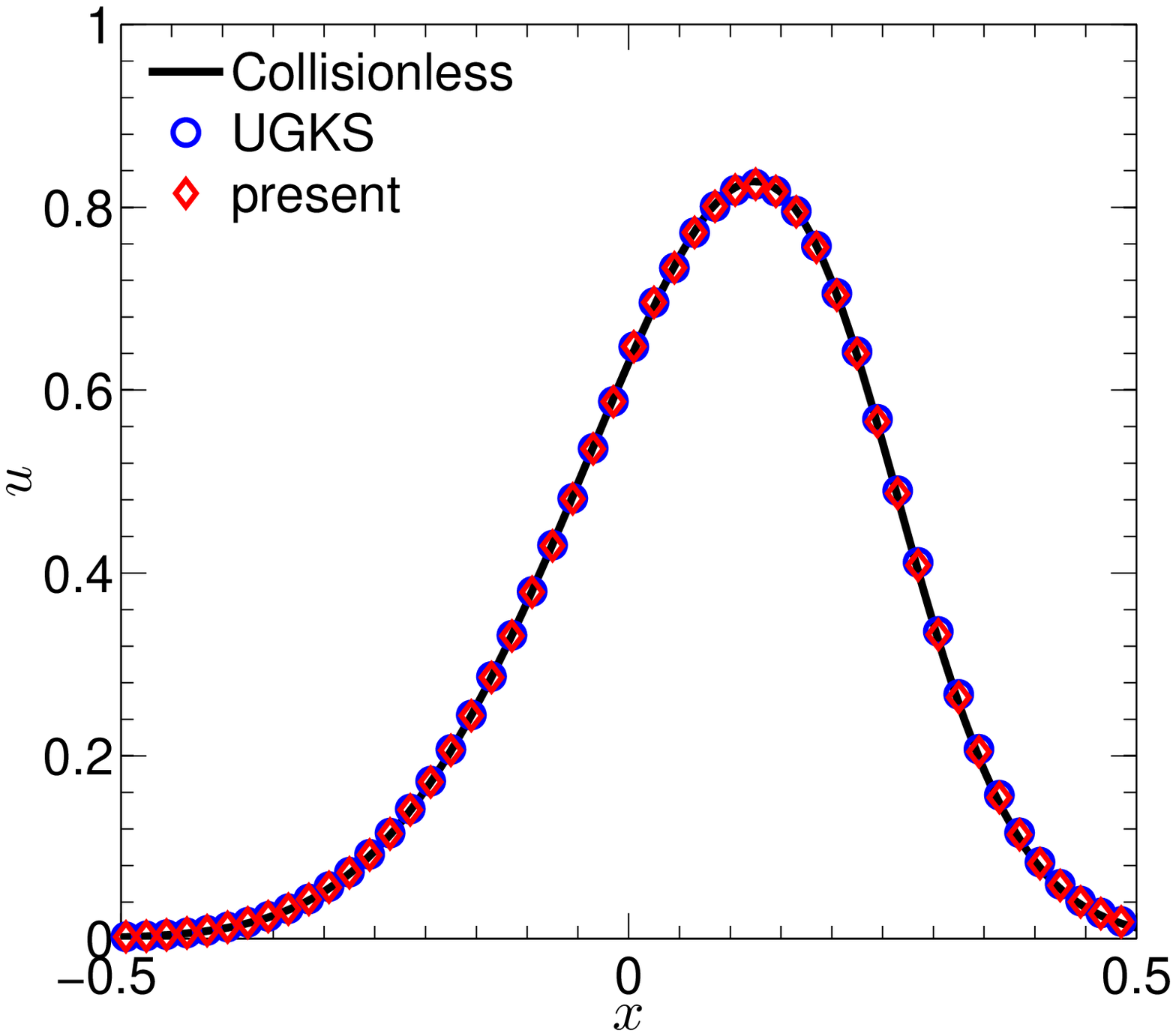}
\caption{(Color online) Density, temperature, and velocity profiles of the shock tube test ($\mu_0=10.0$).}
\label{fig:Tube10}
\end{figure}

\subsection{Shock tube}
The second test case is the standard Sod's shock tube problem \cite{ref:Sod}. The computational domain is $-0.5\le x\le 0.5$, and initially the density, velocity, and pressure are set to be
\begin{equation}
(\rho, u, p)=
\begin{cases}
(\rho_1, u_1, p_1)=(1.0, 0.0, 1.0) & x\le 0;\\
(\rho_2, u_2, p_2)=(0.125, 0.0, 0.1) & x > 0.
\end{cases}
\end{equation}
The gas considered is modeled as hard-sphere molecules such that the viscosity is determined as $\mu=\mu_0(T/T_0)^{0.5}$, where $\mu_0$ is the reference viscosity at reference temperature $T_0$. The reference mean-free-path $\lambda_0$ is then changed by adjusting the $\mu_0$,
\begin{equation}
\lambda_0=\dfrac{16}{5}\dfrac{\mu_0}{p_0}\sqrt{\dfrac{RT_0}{2\pi}},
\end{equation}
where $p_0$ is the reference pressure. Here we take $\rho_1$, $p_1$, and $T_1$ as the reference density, pressure, and temperature, respectively. With different $\mu_0$, the flow will have different degree of rarefaction, which can be used to test the capability of the DUGKS method for simulating flows in different regimes.

In the computation a uniform grid with 100 cells is used to cover the physical domain, and 201 discrete velocities uniformly distributed in $[-10, 10]$ are used to discretize the velocity space,
and the Newton-Cotes quadrature is used to evaluate the velocity moments.
The CFL number is set to be 0.95 in all simulations, and the output time is $t=0.15$.
In all cases the internal freedom is set to be $K=2$ so that the ratio of specific heats is $\gamma=1.4$. In order to make a comparison with the UGKS, $\mu_0$ changes from $10$ to $10^{-5}$ as in Ref. \cite{ref:UGKS}, such that the flow ranges from continuum to free-molecular regimes.

\begin{figure}
\includegraphics[width=0.48\textwidth]{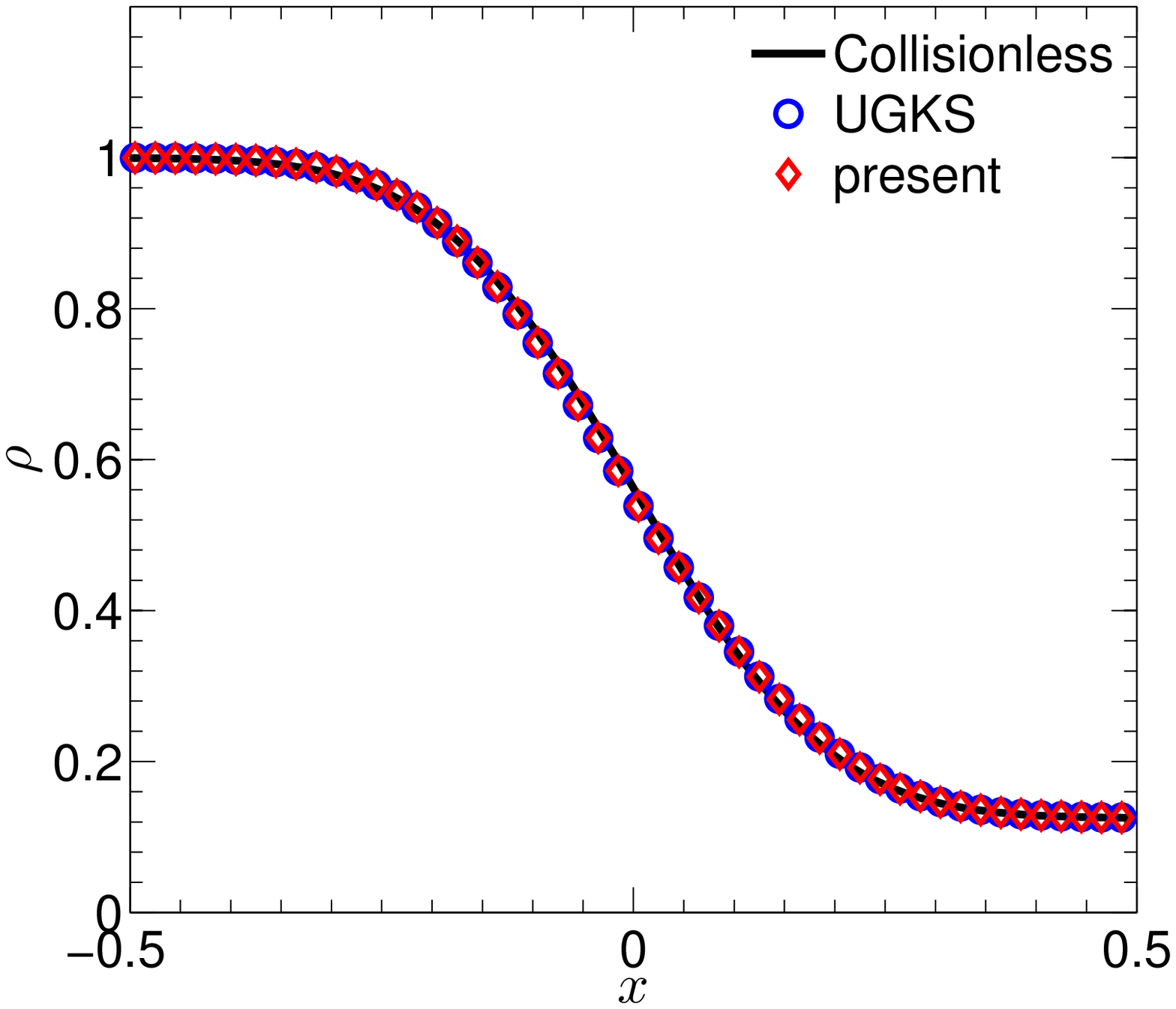}\vfill
\includegraphics[width=0.48\textwidth]{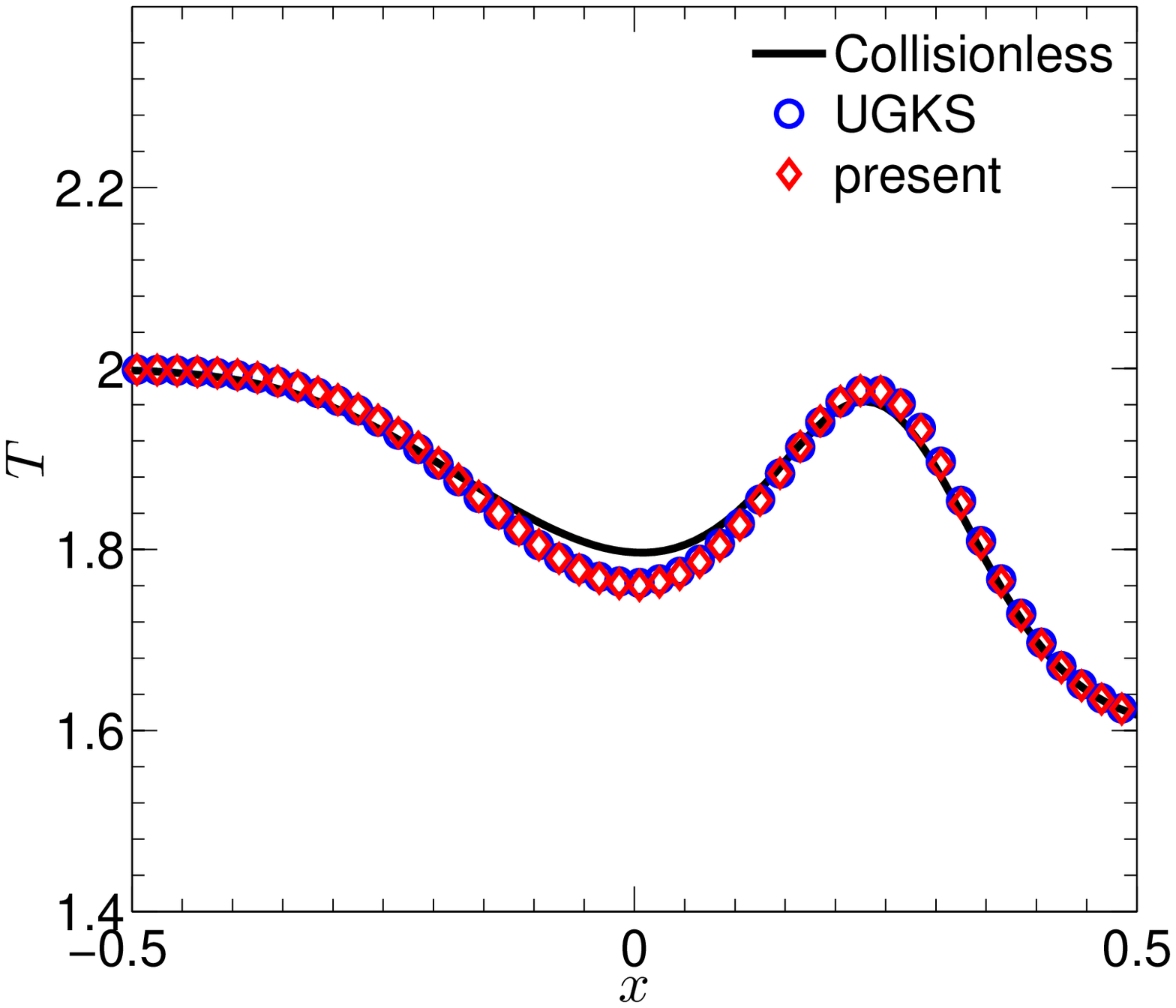}\vfill
\includegraphics[width=0.48\textwidth]{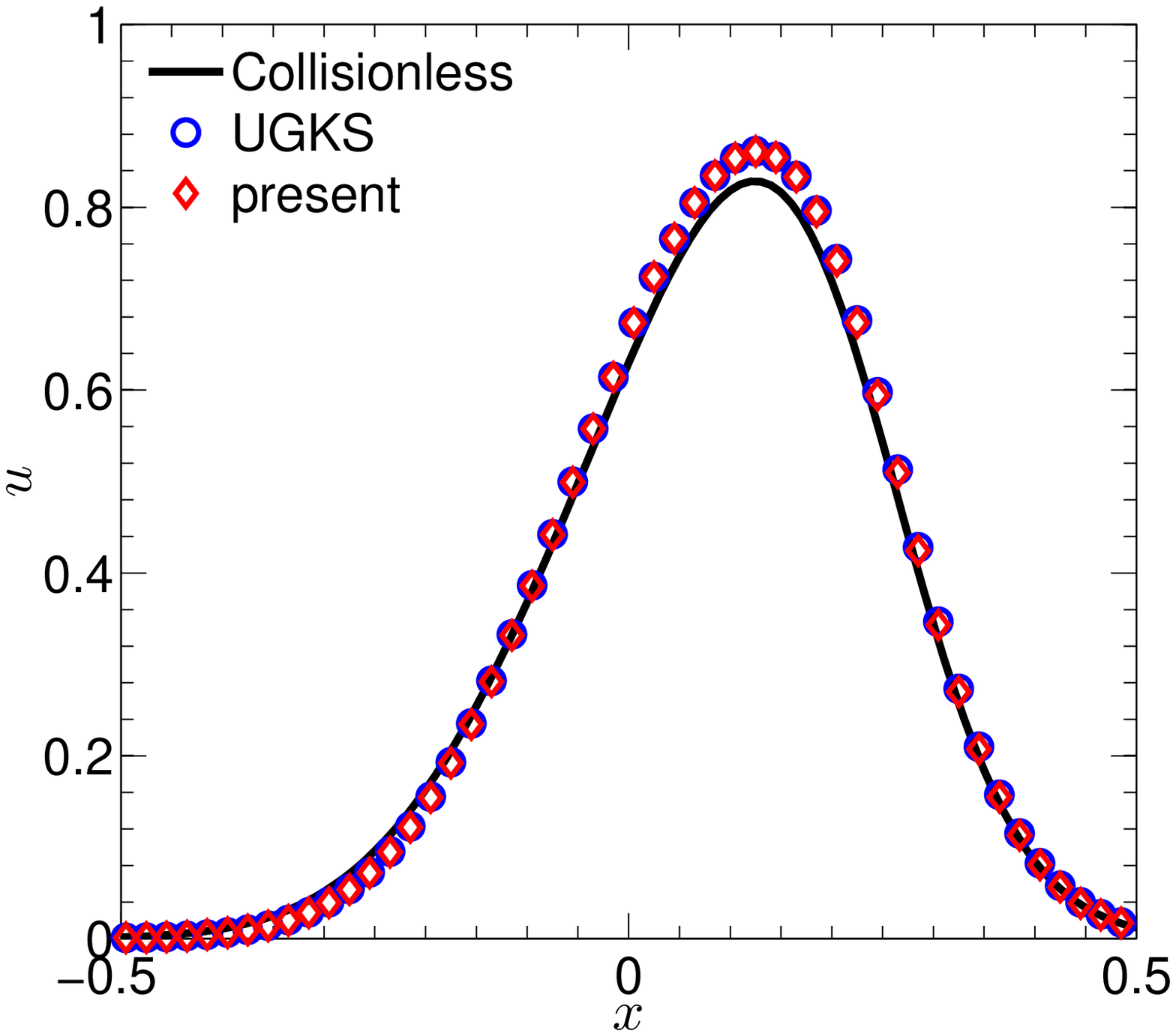}
\caption{(Color online) Density, temperature, and velocity profiles of the shock tube test ($\mu_0=0.1$).}
\label{fig:Tube01}
\end{figure}

Figure \ref{fig:Tube10} shows the density, temperature, and velocity profiles as $\mu_0=10.0$, as well as the UGKS results and the solution of collision-less Boltzmann equation (see Appendix A). In this case the corresponding Knudsen number at the left boundary is about 12.77 and the flow falls in free molecular regime. It can be seen that the DUGKS results agree excellent with the collision-less Boltzmann solution and the UGKS data. As $\mu_0$ decreases to 0.1, the flow falls in the slip regime. The results of the DUGKS in this case is shown in Fig. \ref{fig:Tube01} and compared with the solutions of the UGKS method and collision-less Boltzmann equation. The results of the DUGKS and UGKS are nearly identical and some clear deviations from the collision-less Boltzmann solutions, which is not surprising since collision effects are significant in such case.
\begin{figure}
\includegraphics[width=0.48\textwidth]{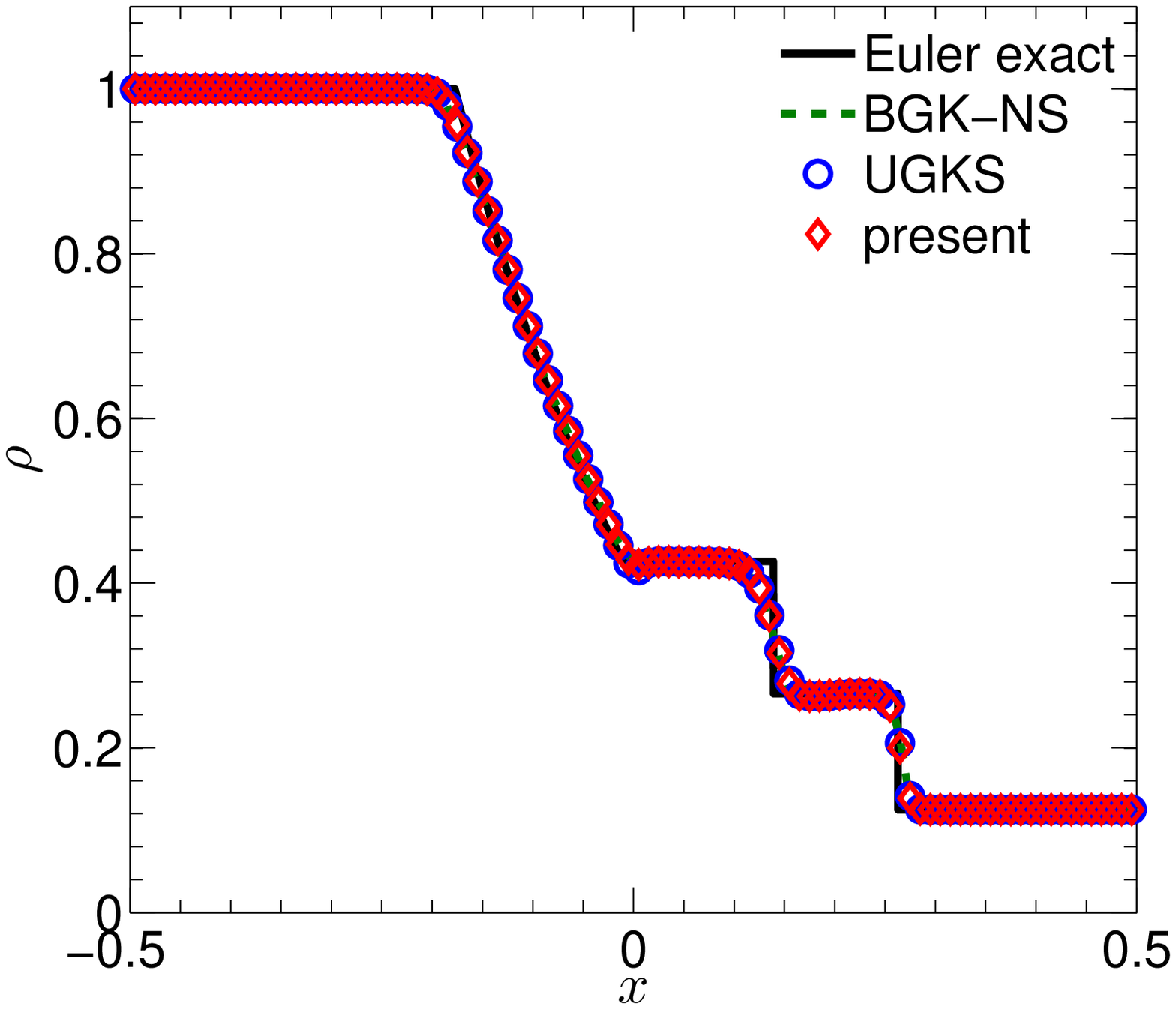}\vfill
\includegraphics[width=0.48\textwidth]{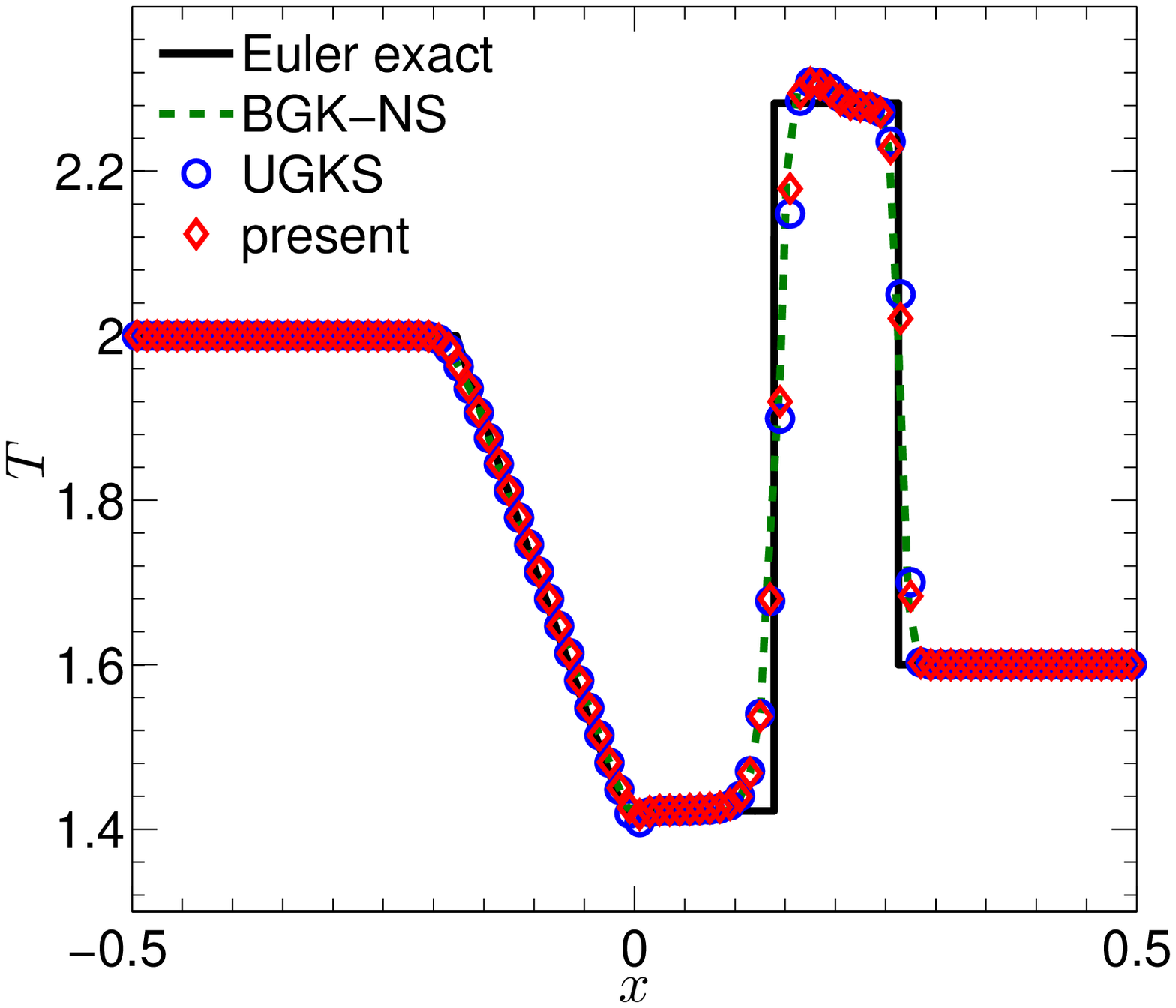}\vfill
\includegraphics[width=0.48\textwidth]{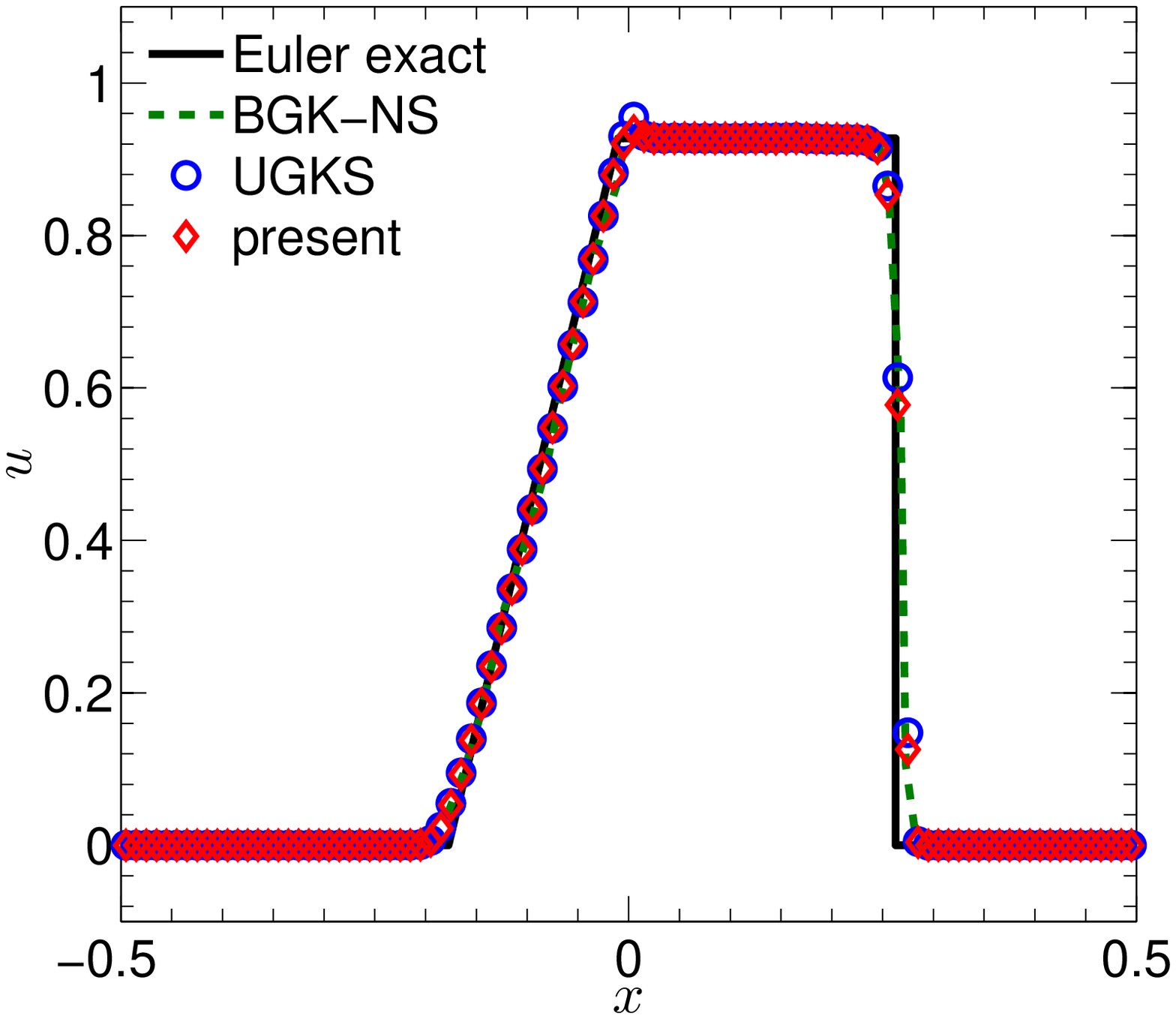}
\caption{(Color online) Density, temperature, and velocity profiles of the shock tube test ($\mu_0=10^{-5}$).}
\label{fig:Tube-5}
\end{figure}

The results for $\mu_0=10^{-5}$ are shown in Fig. \ref{fig:Tube-5}, where the exact solution of the Euler equations,
the results of GKS scheme for Navier-Stokes equations (BGK-NS) \cite{ref:GKS}, and the results of the UGKS scheme, are shown together.
In this case the flow is in the continuum regime and the UGKS becomes a shock capturing scheme for the Euler equations. It can be seen that the DUGKS results agree well with those of the BGK-NS and UGKS methods, but some deviations from the Euler solution are observed.
Particularly, numerical oscillation appears at the contact wave, which may come from the numerical limiter in the reconstruction of flow variables at cell interfaces \cite{ref:UGKS}.

\subsection{Two-dimensional Riemann problem}
We now test the unified property of the DUGKS with the two-dimensional Riemann problem with constant initial data in each quadrant. The solution of the Euler equations for this problem can have a number of different configurations with different initial setups, and a variety of numerical studies have been reported in the past two decades \cite{ref:2DRiemann-Sch93a,ref:2DRiemann-Sch93b,ref:2DRiemann-Zhang90,ref:2DRiemann-Chang95,ref:2D-Riemann-Lax,ref:2D-Riemann-Kurg}. Here we choose one of the typical configurations as listed in Ref. \cite{ref:2D-Riemann-Lax}, where the initial condition is given by
\begin{equation}
(\rho, u, v, p)=
\begin{cases}
(\rho_1, u_1, v_1, p_1)=(0.5313, \ 0,\ 0,\ 0.4), & x>0,\quad y>0,\\
(\rho_2, u_2, v_2, p_2)=(1,\ 0.7276,\ 0,\ 1), & x\le 0,\quad y>0,\\
(\rho_3, u_3, v_3, p_3)=(0.8,\ 0,\ 0,\ 1), & x\le 0,\quad y\le 0,\\
(\rho_4, u_4, v_4, p_4)=(1,\ 0,\ 0.7276, \ 1), & x>0,\quad y\le 0.
\end{cases}
\end{equation}

In our simulations, we set $\mbox{Pr}=2/3$ and $\gamma=1.4$. A $400\times 400$ uniform mesh is employed to discretize the physical domain $0\le x, y\le 1$, and the CFL number is set to be 0.5 in all simulations. As in the one-dimensional shock tube test, a reference viscosity $\mu_0$ at reference temperature $T_0$ is employed to characteristic the rarefication of the gas, and the local viscosity is determined by $\mu=\mu_0 (T/T_0)^w$ with $w=0.5$. At the four boundaries the boundary conditions are set to be $\partial_{\bm{n}}f=0$, where $\bm{n}$ is the outward unit normal vector.

\begin{figure}
\includegraphics[width=0.48\textwidth]{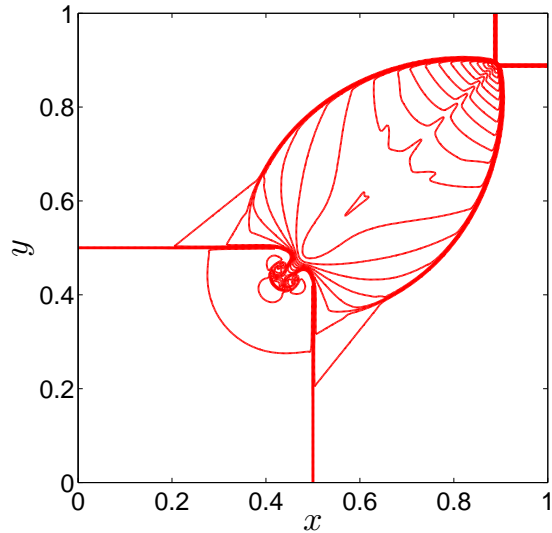}
\caption{(Color online) Density contour of the 2D Riemanm problem ($\mu_0=10^{-7}$).}
\label{fig:Riemann-7}
\end{figure}

We first present the results as $\mu_0=10^{-7}$. The reference mean-free-path $\lambda_0=(\mu_0/p_0)\sqrt{\pi RT_0/2}$
and the collision time $\tau=\mu_0/p_0$ are both in the order of $10^{-7}$, and the flow is in the continuum regime.
In the simulation a $8\times 8$ discrete velocity set based on the half-range Gauss-Hermite quadrature \cite{ref:Shizgal} is employed. The density contour at $t=0.25$ is shown in Fig. \ref{fig:Riemann-7}. It is clear that in this case the DUGKS becomes a shock capture scheme since now $\Delta x=2.5\times 10^{-3}\gg\lambda_0$ and $\Delta t\sim 10^{-4}\gg \tau$. The configuration is also in excellent agreement with the solution of Euler equations by different numerical methods (e.g., \cite{ref:2D-Riemann-Lax,ref:2D-Riemann-Kurg}).

\begin{figure}
\includegraphics[width=0.48\textwidth]{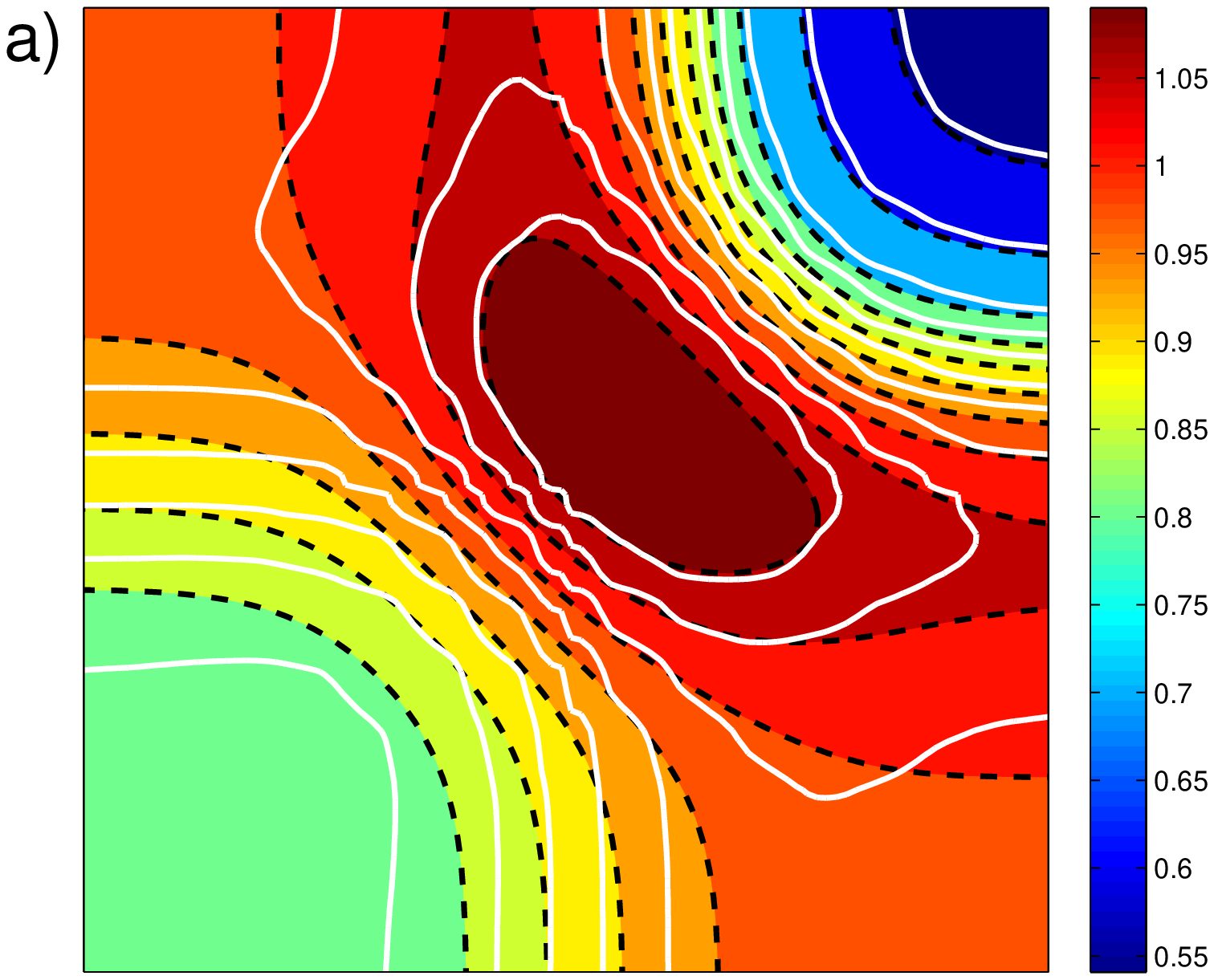}
\includegraphics[width=0.48\textwidth]{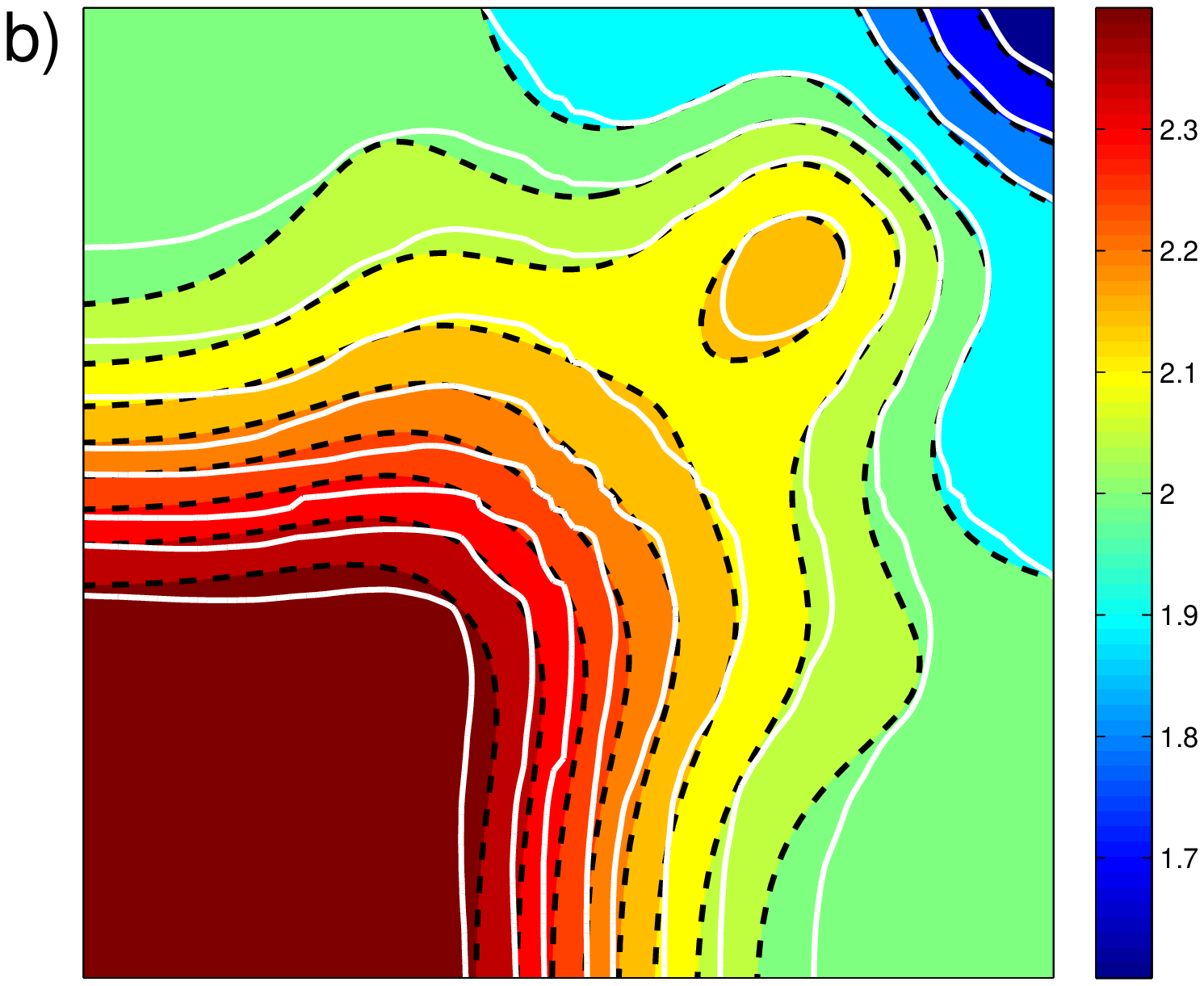}
\includegraphics[width=0.48\textwidth]{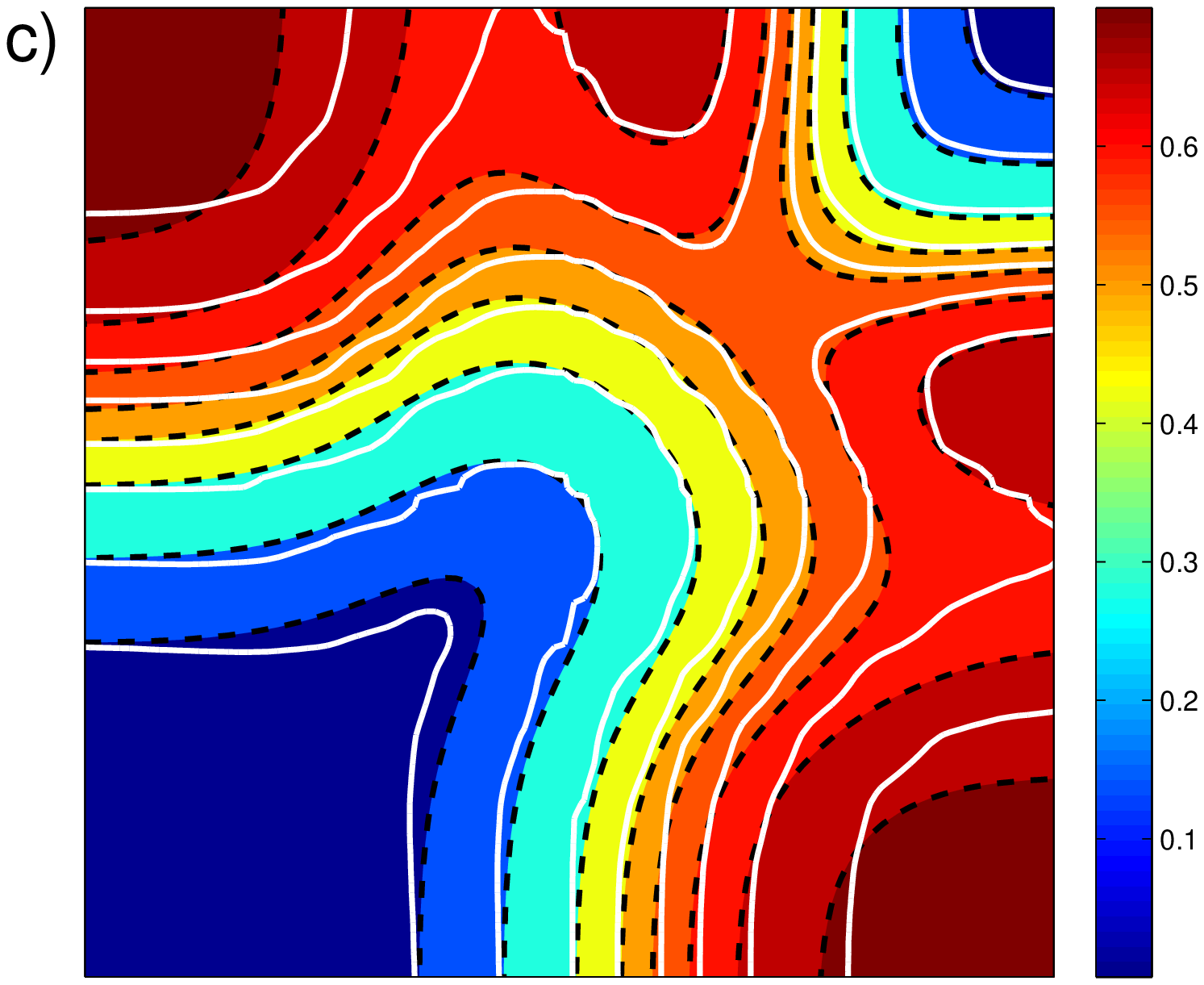}
\includegraphics[width=0.48\textwidth]{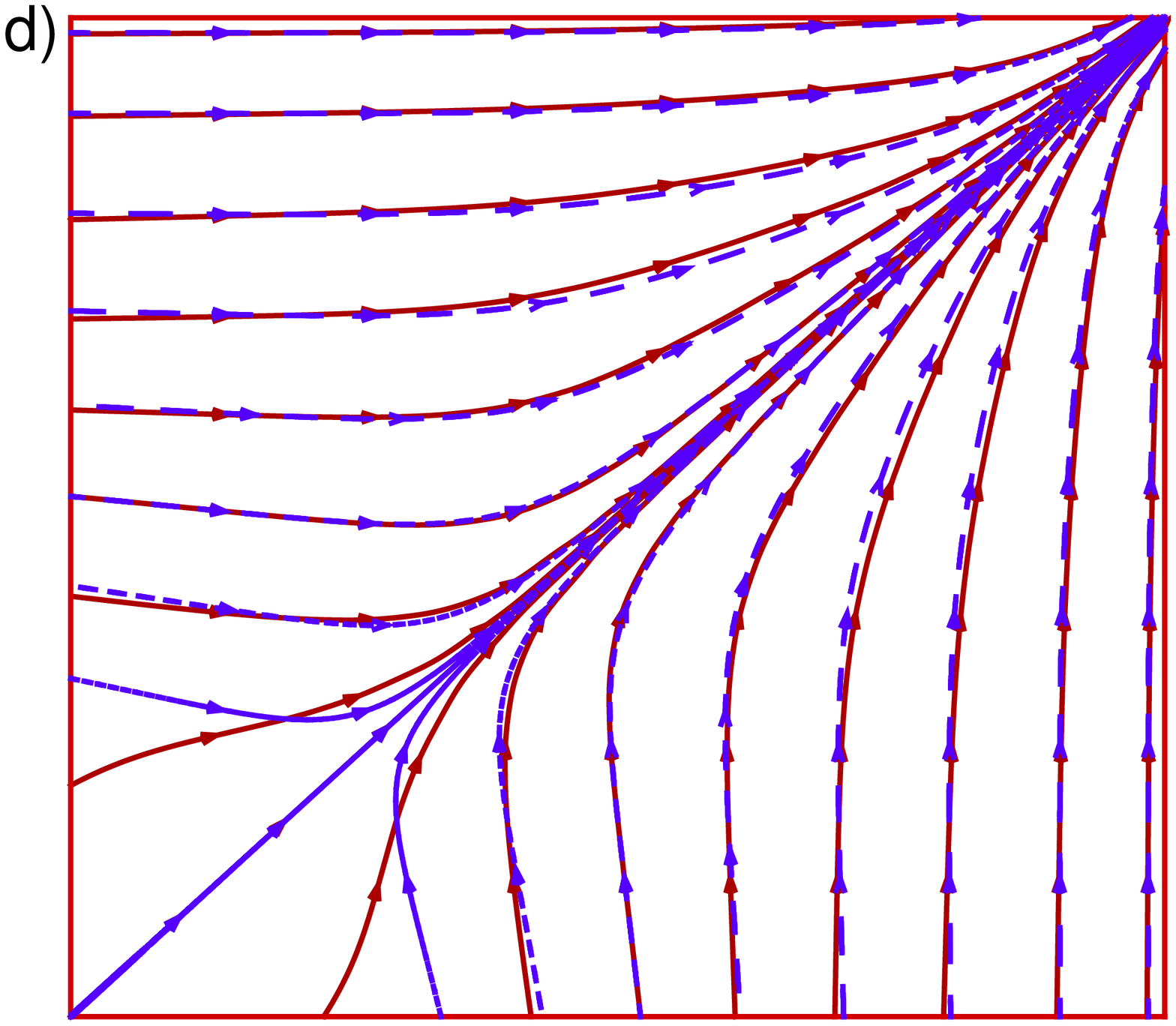}
\caption{(Color online) Contours of the density (a), temperature (b), velocity magnitude (c), and velocity streamlines (d) of the 2D Riemanm problem at $\mu_0=10$. In (a)-(c), the background and dashed lines are from the collision-less Boltzmann equation, and the solid lines are the DUGKS results. In (d), the dashed lines are the solutions of collision-less Boltzmann equation, and the solid lines are the DUGKS results.}
\label{fig:Riemann2D}
\end{figure}

We now test the DUGKS for the problem in free-molecular regime by choosing $\mu_0=10$, respectively. In this case the flow is highly nonequilibrium although the flow field is smooth. In order to capture the nonequilibrium effects, the particle velocity space is discretized with a $201\times 201$ mesh points based on the half-range Gauss-Hermite quadrature \cite{ref:Shizgal}. Furthermore, a uniform mesh with $60\times 60$ cells is used in the physical space which is sufficient to obtain well-resolved solutions. In Fig. \ref{fig:Riemann2D} the density, temperature, velocity magnitude ($(u_x^2+u_y^2)^{1/2}$), and streamlines, are shown at $t=0.15$. For comparison, the results from the solution of collision-less Boltzmann are also presented (see Appendix B). It can be seen that the flow patterns predicted by the DUGKS are quite similar to those of the collision-less Boltzmann equation. The difference may be due to the boundary conditions used in the present simulation where a finite domain is used, while the solutions of the collision-less Boltzmann equation are in the whole infinite domain.
Also, even with $\mu_0 =10$, there is still particle collision in the current DUGKS computation.
However, the overall structure of the two solutions are in good agreement.

The agreement with available data in both continuum and free molecular regimes suggests that the DUGKS has the nice dynamic adaptive property for multi-regime flows,
which is desirable for multiscale flow simulations.

\section{summary}
The multiscale nature of gas flows involving different regimes leads to significant difficulties in numerical simulations.
In this paper a discrete unified gas kinetic scheme in finite-volume formulation is developed for multi-regime flows based on the Shakhov kinetic model.
With the use of discrete characteristic solution of the kinetic equation in the determination of distribution at cell interfaces,
the transport and collision mechanisms are coupled together in the flux evaluation,
which makes the DUGKS a dynamic multiscale approach for flow simulation which distinguishes it from many other numerical methods based on operator splitting approach.
The coupling treatment of transport and collision also makes the DUGKS have some nice features such as multi-dimensional nature and the asymptotic preserving properties.

The DUGKS is validated by several test problems ranging from continuum to free molecular flows.
The numerical results demonstrate the accuracy and robustness of the scheme for multi-regime flow simulations.
The tests also show that the DUGKS exhibits proper dynamic property according to local flow information,
which is important for capturing multiscale flows. In the present simulations, a fixed discrete velocity set is used for each test case.
The computational efficiency can be greatly improved by adopting adaptive velocity techniques \cite{ref:Adaptive_Xu,ref:Adaptive_14}.

\acknowledgments{ZLG acknowledges the support by the National
Natural Science Foundation of China (51125024), and part of the
work was carried out during his visit to the Hong Kong University
of Science and Technology. KX was supported by Hong Kong Research Grant Council
(621011,620813).
}

\appendix
\section{Solution of collision-less Boltzmann equation for the 1D Shock tube problem}
For the shock tube problem, the solution of the collision-less Boltzmann equation is
\begin{equation}
f(x, \xi, t)=
\begin{cases}
f^{eq}(\bm{W}_L), & \xi\ge x/t,\\
f^{eq}(\bm{W}_R), & \xi < x/t.
\end{cases}
\end{equation}
By taking velocity moments of $f(x, \xi, t)$, we can obtain the conserved variables,
\begin{subequations}
\begin{equation}
\rho(x,t)=\dfrac{\rho_1}{2}\erfc(-\tilde{u}_1)+\dfrac{\rho_2}{2}\erfc(\tilde{u}_2),
\end{equation}
\begin{equation}
\rho u(x,t)=\dfrac{\rho_1}{2}\left((2RT_1/\pi)^{1/2} \exp(-\tilde{u}_1^2)+u_1\erfc(-\tilde{u}_1)\right)
+
\dfrac{\rho_2}{2}\left((2RT_2/\pi)^{1/2} \exp(-\tilde{u}_2^2)-u_2\erfc(\tilde{u}_2)\right)
,
\end{equation}
\begin{eqnarray}
\rho E(x,t)&=&\dfrac{\rho_1}{4}\left\{[u_1^2+(K+3)RT_1]\erfc(-\tilde{u}_1)+(u_1+x/t)(2 RT_2/\pi)^{1/2} \exp(-\tilde{u}_2^2)\right\}\nonumber\\
&+&
\dfrac{\rho_2}{4}\left\{[u_2^2+(K+3)RT_2]\erfc(\tilde{u}_2)-(u_2+x/t)(2 RT_2/\pi)^{1/2} \exp(-\tilde{u}_2^2)\right\},
\end{eqnarray}
\end{subequations}
where $\tilde{u}_i=(u_i-x/t)/\sqrt(2\pi R T_i)$, and $\mbox{erfc}$ is the complementary error function defined by
$$
\mbox{erfc}(z)=\dfrac{2}{\sqrt{\pi}}\int_z^{\infty}{e^{-t^2}d t}.
$$

\section{Solution of collision-less Boltzmann equation for the 2D Riemann problem}
For the 2D Riemann problem, the solution of the collision-less Boltzmann equation is
\begin{equation}
f(x, y, \xi_x, \xi_y, \bm{\eta},t)=f^{eq}(x-\xi_x t, y-\xi_y t,\xi_x, \xi_y, \bm{\eta},0).
\end{equation}
Then the conserved variables can be obtained by taking the velocity moments of $f$,
\begin{equation}
\rho(x,y,t)=\dfrac{\rho_1}{4}\erfc(\tilde{u}_1)\erfc(\tilde{v}_1)+\dfrac{\rho_2}{4}\erfc(-\tilde{u}_2)\erfc(\tilde{v}_2)
+\dfrac{\rho_3}{4}\erfc(-\tilde{u}_3)\erfc(-\tilde{v}_3)+\dfrac{\rho_4}{4}\erfc(-\tilde{u}_4)\erfc(-\tilde{v}_4),
\end{equation}
\begin{eqnarray}
\rho u(x,y,t)&=&\dfrac{\rho_1}{4}\left[-\left(2RT_1/\pi\right)^{1/2}e^{-\tilde{u}_1^2}+u_1\erfc(\tilde{u}_1)\right]\erfc(\tilde{v}_1)\nonumber\\
&&+\dfrac{\rho_2}{4}\left[\left({2RT_2}/{\pi}\right)^{1/2}e^{-\tilde{u}_2^2}+u_2\erfc(-\tilde{u}_2)\right]\erfc(\tilde{v}_2)\nonumber\\
&&+\dfrac{\rho_3}{4}\left[\left({2RT_3}/{\pi}\right)^{1/2}e^{-\tilde{u}_3^2}+u_3\erfc(-\tilde{u}_3)\right]\erfc(-\tilde{v}_3)\nonumber\\
&&+\dfrac{\rho_4}{4}\left[-\left({2RT_4}/{\pi}\right)^{1/2}e^{-\tilde{u}_4^2}+u_4\erfc(\tilde{u}_4)\right]\erfc(-\tilde{v}_4),
\end{eqnarray}
\begin{eqnarray}
\rho v(x,y,t)&=&\dfrac{\rho_1}{4}\left[-\left({2RT_1}/{\pi}\right)^{1/2}e^{-\tilde{v}_1^2}+v_1\erfc(\tilde{v}_1)\right]\erfc(\tilde{u}_1)\nonumber\\
&&+\dfrac{\rho_2}{4}\left[-\left({2RT_2}/{\pi}\right)^{1/2}e^{-\tilde{v}_2^2}+v_2\erfc(-\tilde{v}_2)\right]\erfc(-\tilde{u}_2)\nonumber\\
&&+\dfrac{\rho_3}{4}\left[\left({2RT_3}/{\pi}\right)^{1/2}e^{-\tilde{v}_3^2}+v_3\erfc(-\tilde{v}_3)\right]\erfc(-\tilde{u}_3)\nonumber\\
&&+\dfrac{\rho_4}{4}\left[-\left({2RT_4}/{\pi}\right)^{1/2}e^{-\tilde{v}_4^2}+v_4\erfc(\tilde{v}_4)\right]\erfc(\tilde{u}_4),
\end{eqnarray}
and
\begin{subequations}
\begin{equation}
\rho E(x,y,t)=\dfrac{1}{8}\left(\rho_1 J_1+\rho_2 J_2+\rho_3 J_3+\rho_4 J_4\right),
\end{equation}
with
\begin{eqnarray}
J_1(x,y,t)&=&-\left[\left({y}/{t}+v_1\right) e^{-\tilde{v}_1^2}\erfc(\tilde{u}_1)
+\left({x}/{t}+u_1\right)e^{-\tilde{u}_1^2}\erfc(\tilde{v}_1)\right]\left({2RT_1}/{\pi}\right)^{1/2}\nonumber\\
&&+\left[(K+2)RT_1+u_1^2+v_1^2\right]\erfc(\tilde{u}_1)\erfc(\tilde{v}_1),
\end{eqnarray}
\begin{eqnarray}
J_2(x,y,t)&=&-\left[\left({y}/{t}+v_2\right) e^{-\tilde{v}_2^2}\erfc(-\tilde{u}_2)
-\left({x}/{t}+u_2\right) e^{-\tilde{u}_2^2}\erfc(\tilde{v}_2)\right]\left({2RT_2}/{\pi}\right)^{1/2}\nonumber\\
&&+\left[(K+2)RT_2+u_2^2+v_2^2\right]\erfc(-\tilde{u}_2)\erfc(\tilde{v}_2),
\end{eqnarray}
\begin{eqnarray}
J_3(x,y,t)&=&\left[\left({y}/{t}+v_2\right)e^{-\tilde{v}_3^2}\erfc(-\tilde{u}_3)
+\left({x}/{t}+u_3\right)e^{-\tilde{u}_3^2}\erfc(-\tilde{v}_3)\right]\left({2RT_3}/{\pi}\right)^{1/2}\nonumber\\
&&+\left[(K+2)RT_3+u_3^2+v_3^2\right]\erfc(-\tilde{u}_3)\erfc(-\tilde{v}_3),
\end{eqnarray}
\begin{eqnarray}
J_4(x,y,t)&=&\left[\left({y}/{t}+v_4\right)e^{-\tilde{v}_4^2}\erfc(\tilde{u}_4)
-\left({x}/{t}+u_4\right)e^{-\tilde{u}_4^2}\erfc(-\tilde{v}_4)\right]\left({2RT_4}/{\pi}\right)^{1/2}\nonumber\\
&&+\left[(K+2)RT_4+u_4^2+v_4^2\right]\erfc(\tilde{u}_4)\erfc(-\tilde{v}_4),
\end{eqnarray}
\end{subequations}
where $\tilde{\bm{u}}_i=(\bm{u}_i-\bm{x}/t)/\sqrt{2RT_i}$.

\end{document}